\documentclass[acmtog,anonymous=false,review=false]{acmart}

\usepackage{cirl}
\usepackage{geometrycollective}
\usepackage{WoStDiff}

\setcitestyle{square,nosort}

\setcopyright{rightsretained}
\acmJournal{TOG}
\acmYear{2024}
\acmVolume{43}
\acmNumber{6} 
\acmArticle{174}
\acmMonth{12}
\acmDOI{10.1145/3687913}

\begin{teaserfigure}
   \centering
   \includegraphics{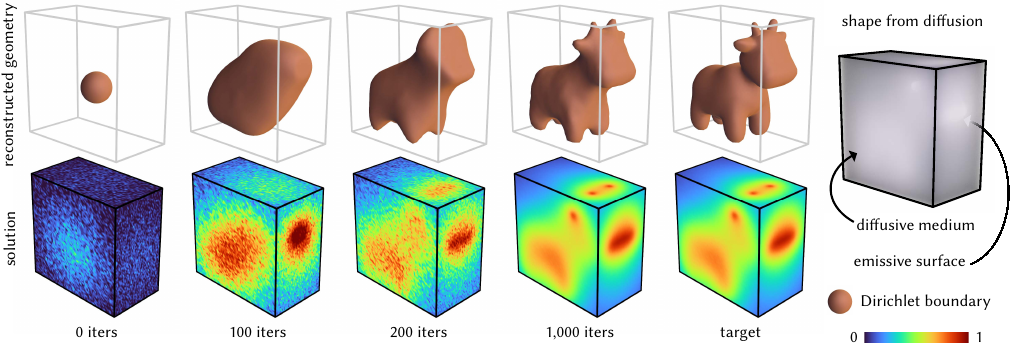}
    \caption{
       For a given boundary value problem, our differential walk on spheres algorithm makes it possible to differentiate solution values with respect to problem parameters.  Here we consider an inverse problem where we recover the shape of an emissive object from its observed diffusion profile on the boundary of a box, via gradient-based optimization. Unlike conventional mesh- or grid-based approaches, we can evaluate derivatives at points of interest without needing to compute a global solution (here, only at the observed points).
    \label{fig:teaser}}
\end{teaserfigure}

\begin{document}
\title{Differential Walk on Spheres}

\author{Bailey Miller}
\email{bmmiller@andrew.cmu.edu}
\affiliation{%
  \institution{Carnegie Mellon University}
  \streetaddress{5000 Forbes Ave}
  \city{Pittsburgh}
  \state{PA}
  \postcode{15213}
  \country{USA}
}
\orcid{0009-0009-0881-0351}

\author{Rohan Sawhney}
\email{rsawhney@nvidia.com}
\affiliation{
  \institution{NVIDIA}
  \streetaddress{2788 San Tomas Expy}
  \city{Santa Clara}
  \state{CA}
  \postcode{95051}
  \country{USA}
}
\orcid{0000-0002-3661-1554}

\author{Keenan Crane}
\email{kmcrane@cs.cmu.edu}
\affiliation{
  \institution{Carnegie Mellon University}
  \streetaddress{5000 Forbes Ave}
  \city{Pittsburgh}
  \state{PA}
  \postcode{15213}
  \country{USA}
}
\orcid{0000-0003-2772-7034}

\author{Ioannis Gkioulekas}
\email{igkioule@cs.cmu.edu}
\affiliation{%
  \institution{Carnegie Mellon University}
  \streetaddress{5000 Forbes Ave}
  \city{Pittsburgh}
  \state{PA}
  \postcode{15213}
  \country{USA}
}
\orcid{0000-0001-6932-4642}

\begin{abstract}
We introduce a Monte Carlo method for computing derivatives of the solution to a partial differential equation (PDE) with respect to problem parameters (such as domain geometry or boundary conditions). Derivatives can be evaluated at arbitrary points, without performing a global solve or constructing a volumetric grid or mesh. The method is hence well suited to inverse problems with complex geometry, such as PDE-constrained shape optimization. Like other \emph{walk on spheres (WoS)} algorithms, our method is trivial to parallelize, and is agnostic to boundary representation (meshes, splines, implicit surfaces, \etc{}), supporting large topological changes.  We focus in particular on screened Poisson equations, which model diverse problems from scientific and geometric computing.  As in differentiable rendering, we jointly estimate derivatives with respect to all parameters---hence, cost does not grow significantly with parameter count.  In practice, even noisy derivative estimates exhibit fast, stable convergence for stochastic gradient-based optimization, as we show through examples from thermal design, shape from diffusion, and computer graphics.
\end{abstract}

%

%
%

\maketitle

\section{Introduction}\label{sec:intro}

Which shape best explains observed physical behavior?  How can one design shapes that maximize (or minimize) a target physical quantity?  Such \emph{inverse problems} are fundamental to numerous challenges in science and engineering.  For instance, one might need to assess damage to an airplane wing using indirect thermal measurements~\cite{zalameda2014thermal}, or infer the shape of a tumor through deep layers of tissue \citep{arridge1999optical}.  Likewise, one might seek to design circuit geometry that maximizes dissipation of heat \citep{Zhan:2008:ThermalDesign}, airfoils that generate prescribed lift \citep{Hicks:1977:WingDesign}, or lightweight structures that withstand significant load \citep{allaire2014shape}.  To solve such problems, one must be able to efficiently and accurately differentiate solutions to partial differential equations (PDEs) with respect to the shape of the domain, or its boundary conditions \citep{hadamard1908memoire,cea1973quelques}.  However, for problems with complex geometry, even just solving such PDEs on a fixed domain can be daunting for traditional methods---making many important inverse problems unapproachable.

\begin{figure}[t]
	\centering
	\includegraphics[width=\columnwidth]{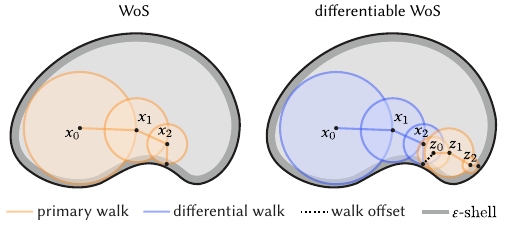}
	\caption{\figloc{Left:} Walk on spheres (\cref{alg:wos}) recursively jumps to a random point on the largest sphere around the current walk location. The walk is terminated when it reaches the $\varepsilon$-shell, where the Dirichlet condition is evaluated at the closest point on the boundary. \figloc{Right:} \Diffbvp{} WoS (\cref{alg:wos_deriv}) also takes a walk to the $\varepsilon$-shell, but additionally launches a \forwardbvp{} walk from an offset point close to the boundary to estimate the \diffbvp{} boundary condition.}
	\label{fig:coupling}
\end{figure}

The \emph{walk on spheres (WoS)} method \citep{Muller:1956:WOS,Sawhney:2020:MCG} and its recent extensions \citep{svWoS, WoSt, WoStRobin} provide \emph{grid-free} alternatives to traditional PDE solvers that entirely bypass the need for volumetric mesh generation or global solves. Compared to finite element or finite difference techniques, these Monte Carlo-based methods enable \emph{pointwise} evaluation of the solution to linear elliptic PDEs such as the Poisson equation, through simulation of independent random walks (\cref{fig:coupling}, \emph{left}). As explored by \citet{Sawhney:2020:MCG} and in followup work \citep{svWoS, WoSt, WoStRobin}, this approach offers a degree of geometric robustness and flexibility that has previously eluded traditional solvers. Yet WoS has historically been limited to \emph{forward} tasks (\ie, solving PDEs on a given domain), rather than inverse tasks such as geometric optimization.

How can we adapt WoS to inverse problems?  One idea is to apply \emph{automatic differentiation} to existing algorithms---but as in differentiable rendering this na\"{i}ve strategy is undesirable for several reasons:
\begin{enumerate*}
	\item Automatic differentiation, when na\"{i}vely applied to algorithms like WoS and Monte Carlo path tracing, computes derivatives for accelerated geometric queries which is both inefficient and impractical \citep{niemeyer2020differentiable}.
	\item It results in exponential memory complexity and quadratic computational complexity, compared to the linear memory and computational complexity of the recursive forward algorithms \citep{nimier2020radiative,vicini2021path}.
	\item It cannot compute correct derivatives with respect to the evolving geometry that geometric optimization and inverse problems typically require \citep{Li:2018:EdgeSampling,Loubet:2019:Reparameterizing,Bangaru:2020:WarpedArea,Zhang:2020:PSDR}.
	\item Even in constrained settings where these critical limitations may not apply, automatic differentiation of Monte Carlo algorithms results in differential estimators with suboptimal statistical performance \citep{zeltner2021monte}.
\end{enumerate*}

We instead take a more principled approach, and formulate PDEs that compute derivatives (\aka{} \emph{sensitivities}) with respect to a collection of parameters \(\param\).  Crucially, these PDEs can again be estimated via WoS algorithms---which must now be adapted to handle nesting between the PDEs characterizing the derivatives, and the solution itself (\cref{sec:bvp_deriv}).  The parameters $\param$ are analogous to the \emph{scene parameters} in differentiable rendering---for example, they might describe the vertex locations of a triangular mesh, the anchors and tangents of a cubic spline, the centers of radial basis functions defining an implicit surface, or the pose of a rigid object.  Our method simultaneously estimates derivatives for all parameters using a set of nested random walks, making it feasible to optimize systems with thousands of degrees of freedom (\eg{}, a densely-sampled mesh).

The Monte Carlo approach is also well-matched to the needs of descent-based optimization.  Far from local minima, the gradient is merely a heuristic for the ideal descent direction---making exact gradient computation (as in traditional methods) overkill in most scenarios.  A stochastic approach enables us to reduce the computational load in early stages of optimization by taking fewer samples, increasing sample count only as we approach a solution.  In fact, as in differentiable rendering \citep{Gkioulekas2013:IVR,Li:2018:EdgeSampling,Zhang:2020:PSDR,nimier2020radiative,vicini2022differentiable,zhaocourse}, noisy derivatives can actually \emph{improve} optimization quality by helping to avoid local minima (\cref{sec:sgd}).  Moreover, our derivatives do not suffer from numerical robustness issues that can plague traditional PDE solvers, such as exploding derivatives due to ill-conditioned elements on an evolving mesh.

\paragraph{Scope} Although our approach applies in principle to many of the PDEs handled by current (forward) WoS algorithms (\eg{}, with drift terms or variable coefficients), we choose to focus this paper on \emph{screened Poisson equations}, which strike a good balance between generality and conceptual and expositional simplicity. Such equations find widespread use in geometry processing, simulation, graphics, and scientific computing. For example, they can model the steady-state temperature inside a conductive solid, the isotropic distribution of light deep inside a scattering medium, or the diffusive interpolation of boundary values in an inpainted image. We show geometric optimization examples inspired from these settings in \cref{sec:geometric_examples}, and choose these examples to also showcase the ability of our approach to optimize complex and varied geometric representations---triangular meshes, B\'{e}zier curves, and implicit surfaces. We provide a video visualizing these optimization examples, and an open-source implementation of our solver, on the project website.%
\footnote{{\url{https://imaging.cs.cmu.edu/differential_walk_on_spheres}}}

\section{Related Work}\label{sec:related}

Optimizing the parameters of a system, such as its shape, to improve performance or meet a desired objective is a widely studied mathematical and algorithmic topic. Here we broadly review prior work on inverse design problems that involve solving PDEs, followed by Monte Carlo methods for differentiable simulation in graphics.

\subsection{PDE-constrained shape optimization}\label{sec:pde-shape-opt-related}

Many application domains require studying how the shape of a system evolves under small changes to a PDE solution---for the transfer of heat \citep{zhuang2007level,feppon2019shape}, deformation of elastic bodies \cite{ameur2004level,allaire2014shape}, flow of fluids \citep{Treuille:2003:Smoke}, distortion of mappings \citep{Sharp:2018:VSC}, interpolation of colors \citep{zhao2017inverse}, \etc{} The mathematical roots of PDE-constrained shape optimization can be traced back to the method of Hadamard \citep{hadamard1908memoire}, which computes the sensitivity of a problem subject to small deformations of its boundary. With the advent of powerful computers and advanced numerical methods like the finite element method (FEM), the mathematical framework of shape optimization was further developed by authors such as C{\'e}a \citep{cea1973quelques,cea1974adaptation}, wherein techniques such as implicit differentiation and the adjoint method became popular for efficiently computing derivatives of objective functions dependent on PDE solutions \citep{henrot2018shape}. We use these classical results, specifically for the screened Poisson equation as reviewed in \cref{sec:optimization}, as the starting point for developing our differential Monte Carlo method in \cref{sec:derivatives}.  Note that \citet{zhao2017inverse} and \citet[Section 5.8]{henrot2018shape} show how to express shape-functional derivatives in terms of the solution to an \emph{adjoint boundary value problem}---we compare this formulation to ours in \cref{app:adjoint}.

\subsection{Algorithms for shape optimization}\label{sec:diff-sim-algos-related}

The earliest numerical methods for shape optimization were based on evolving meshes \citep{Pironneau:1984:OptimalShapeDesign,allaire:2006:structural}, where a discretized domain serving as an initial guess is deformed under PDE constraints to better match an objective. Over the past two decades, level-set based methods \citep{ameur2004level,li2005level,zhuang2007level,allaire2014shape,feppon2019shape,lebbe2019robust,gropp2020implicit} have since superseded purely mesh-based approaches---here a shape described implicitly as a level set function is evolved on a fixed background mesh or grid. This representation enables complex topological changes to the optimized shape (\eg, merging of holes, region splitting) without needing to remesh, which is particularly desirable for optimizations where the final shape is unknown. Hybrid approaches have also been developed, allowing for both topological changes and more accurate PDE solutions via conforming mesh boundaries \citep{allaire2013mesh}. More recently, \emph{physics-informed neural networks} have emerged as alternatives to traditional PDE techniques \citep{raissi2019physics}; these methods can face difficulties in enforcing hard PDE constraints in complex geometric domains \citep{krishnapriyan2021characterizing}, and are currently in their infancy for shape optimization tasks \citep{lu2021physics}.

\paragraph{Grid-based methods in graphics} Computer graphics has developed specialized algorithms for shape optimization, to handle the challenging design of fluidic systems~\citep{Du:2020:StokesOptimization, li2022fluidic}, rigid structures~\citep{Zhu:2017:TopologyOptimization, Liu:2018:TopologyOptimization, Whiting2012Structural}, and acoustic filters~\citep{Li:2016:AcousticVoxels}. These methods build on decades of shape optimization research that leverages grid-based techniques \citep{feppon2019shape}. Though practical for their intended applications, these methods are designed for a specific geometric representation and require global solves to compute gradients during optimization. 

By contrast, our method can work directly with a variety of boundary representations, such as meshes, splines, NURBS \citep{Marschner:2021:SumOfSquares}, and even implicit surfaces \citep{sharp2022spelunking,Gillespie:2024:RTH} that allow for large topological changes (\cref{sec:geometric_examples}), without the need for any specialization---the only requirement is to be able to query the distance to the boundary. Another key benefit of our Monte Carlo approach over grid-based alternatives is \emph{output-sensitive optimization}, \ie, the ability to evaluate derivatives pointwise in regions of interest without a \emph{global} PDE solve (\cref{fig:toaster}). Our method also avoids the many challenges associated with grid generation and adaptive refinement---for instance, it does not have to contend with numerical issues faced by mesh-based approaches from ill-conditioned elements. Moreover, even though derivative estimates computed with our method contain noise, they are correct in expectation. We can therefore improve their accuracy at any stage of the optimization by simply taking more samples, in place of, \eg, increasing the resolution of a background grid.

\subsection{Differentiable Monte Carlo simulation}\label{sec:diff-mc-sim-related}
Stochastic gradient optimization has had a profound impact on machine learning, enabling efficient training of complex models on large datasets without overfitting \citep{kingma2015adam}. In recent years, it has also become popular for differentiable physics simulation \citep{Nimier2019Mitsuba,Hu:2020:DiffTaichi,warp2022} and geometric optimization \citep{remelli2020meshsdf,shen2021deep,Shen:2023:Flexicube} in computer graphics. For instance, it is the method of choice for differentiable Monte Carlo rendering \citep{Gkioulekas2013:IVR,Li:2018:EdgeSampling,Zhang:2020:PSDR,nimier2020radiative,vicini2022differentiable,zhaocourse}, where additional stochasticity in the derivative estimates of scene parameters (\eg, lighting, materials and geometry) has been observed to act as a regularizer for highly underdetermined inverse rendering problems. When paired with 3D geometry representations that allow for easy topological changes, differentiable Monte Carlo rendering reliably recovers a target geometry using a random or uninformed initialization of scene parameters \citep{Hasselgren:2022:Shape, Cai:2022:PSDR-MeshSDF, Ishit:2022:LevelSetTheory}. Our method likewise employs stochastic optimization with noisy derivative estimates for PDE-constrained shape optimization and may be paired with a range of boundary representations. Similar to differentiable rendering, cost does not increase significantly with the number of parameters, as during optimization we simultaneously estimate derivatives with respect to all parameters with a single PDE solve.

\begin{figure}[t]
	\centering
	\includegraphics[width=\columnwidth]{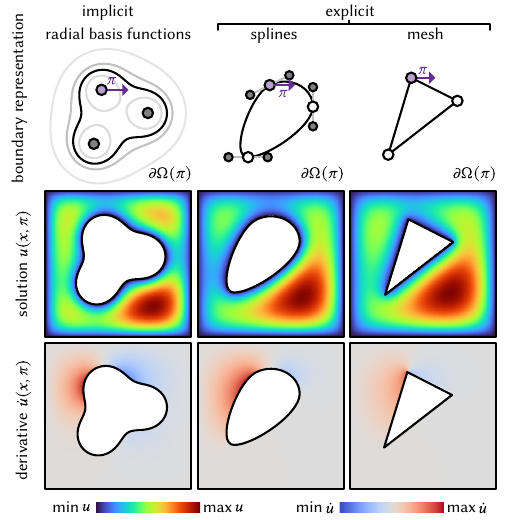}
	\caption{We can compute derivatives on a wide range of boundary representations, including those not directly handled by conventional solvers: implicit surfaces (\figloc{left}) and splines (\figloc{center}). For a solution to a Poisson equation $\solution\paren{\point, \param}$ (\figloc{middle row}), here we show the derivatives $\paramderiv{\solution}\paren{\point, \param}$ (\figloc{bottom row}) with respect to positional parameters $\param$ of the boundary $\boundary\paren{\param}$.}
	\label{fig:forward_derivatives}
\end{figure}

\subsection{Walk on spheres}\label{sec:walk-on-spheres-related}

We develop our differential Monte Carlo method using the walk on spheres algorithm \citep{Muller:1956:WOS}, which has become a popular grid-free alternative to traditional PDE solvers in the graphics community due to its striking similarities with Monte Carlo ray tracing \citep{Sawhney:2020:MCG}. In a short span of time, WoS has been generalized to solve a much broader class of linear diffusive PDEs \citep{nabizadeh2021kelvin,svWoS,WoSt,WoStRobin,Lambilly:2023:Heat}; a number of variance reduction techniques have also been developed to improve estimation quality \citep{qi2022bidirectional,BVC,bakbouk2023mean,li2023neural}, along with explorations into more advanced applications such as thermal imaging \citep{bati2023coupling} and fluid simulation \citep{Rioux:2022:MCFluid,Jain:2024:MCFluid,sugimoto2024velocity}. Though we focus primarily on screened Poisson equations with Dirichlet boundary conditions, our method can in principle be generalized to handle the larger set of equations WoS can solve, \eg, we describe an extension to mixed Dirichlet-Robin boundary conditions in \cref{sec:diffwost}. Moreover, as our approach essentially boils down to coupling two random walks simulated using WoS, it should benefit directly from any future improvements to WoS estimators.

\paragraph{WoS for inverse problems.} \citet{Yilmazer:2022:dWoS} considered a differential version of WoS for optimizing parameterized source terms and PDE coefficients on a \emph{fixed} domain. Our emphasis is on optimizing the shape of the domain and the boundary conditions imposed on it. These differentiable simulation capabilities are complementary and useful for different types of inverse problems.

\paragraph{Walk on boundary alternative.}\label{sec:walk-on-boundary}
The walk on boundary method \citep{sugimoto2023wob, sabelfeld2013rwb} is another Monte Carlo PDE solver that performs random walks over the boundary of a domain via ray intersection queries. Designing a differential walk on boundary method is conceptually attractive given the close resemblance of the forward process to Monte Carlo rendering; however, prior work \citep{WoSt, WoStRobin} demonstrates that the practical application of walk on boundary is hindered by the lack of convergence guarantees. In particular, in the presence of either non-convex geometry or mixed boundary conditions, walk on boundary estimates can fail to converge. By contrast, walk on spheres is provably convergent and its practical application to complex domains is well demonstrated by prior works \citep{Sawhney:2020:MCG, WoSt, WoStRobin}.

\paragraph{Concurrent work.} In concurrent work, \citet{Yu:2024:DiffPoisson} develop a similar algorithm to ours for shape optimization constrained by the screened Poisson equation. Our work differs in a few key ways:
\begin{enumerate*}
	\item The theoretical framework we develop our algorithm on translates the problem of computing derivatives to one of solving boundary value problems (BVPs) of the same type as the PDE in consideration (\cref{sec:optimization}). This framework allows us to easily extend our approach to more general BVPs (Sections \ref{sec:robin} \& \ref{sec:diffwost}) using existing estimators that generalize WoS \citep{WoSt,WoStRobin}, which we demonstrate in a thermal design problem that includes mixed Dirichlet-Robin boundary conditions (\cref{sec:thermal_design}). In principle, our approach also enables extending other existing estimators for this PDE (\eg, bidirectional or reverse random walks \citep{qi2022bidirectional,WoStRobin}) to the differential setting.
	\item We provide an unbiased and efficient approach for calculating reverse-mode derivatives of shape functionals using \emph{U-statistics} (\cref{sec:backward}).
	\item We show more varied and challenging shape optimization applications, including optimization of complex 3D meshes, implicit geometry, and B\'{e}zier curves, to help showcase the benefits of Monte Carlo techniques for PDE-constrained shape optimization. 
\end{enumerate*}

\section{Background}\label{sec:background}
We first review concepts from surface evolution and PDE theory, as well as the walk on spheres algorithm.  For in-depth discussion of differential surface evolution, we refer to \citet{henrot2018shape}.

\subsection{Surface evolution}\label{sec:continuum}
Throughout we optimize a domain \(\domain\paren{\param} \subset \R^3\) with boundary \(\boundary\paren{\param}\), whose geometry is encoded by a finite set of parameters \(\param \in \R^N\).  (\cref{sec:geometric_examples} details several possibilities.) For any $\pi$-dependent function \(\phi\), we denote by \(\nabla\phi\) partial differentiation with respect to its domain (\ie, the \emph{spatial gradient}); and by \(\paramderiv{\phi}\) partial differentiation with respect to \(\param\) (\ie, the \emph{parameter derivative}).

We assume that the parameterization of \(\boundary\paren{\param}\) is differentiable, so that an infinitesimal perturbation of \(\param\) corresponds to an infinitesimal deformation of \(\boundary(\param)\), described by a \emph{velocity field} \(\dispField: \boundary\paren{\param} \to \R^3\).  Since tangential motion does not change the geometry of \(\boundary\paren{\param}\), we often require only the \emph{normal velocity} \(\normalDispField \equiv \normal \cdot\dispField\), where \(\normal: \boundary\paren{\param} \to \R^3\) is the outward unit normal field.  Expressions for the normal velocity depend on whether we model $\boundary\paren{\param}$ as an implicit or explicit surface (\cref{fig:forward_derivatives}).

\subsubsection{Implicit surface}
\label{sec:ImplicitSurfaceRepresentation}
In this case, we model the boundary $\boundary\paren{\param}$ as the zero-level set of an \emph{implicit function} \(\implicit\paren{\cdot, \param}: \R^3 \to \R\) controlled by \(\param\); that is,  $\boundary\paren{\param} \coloneqq \curly{\point \in \R^3 : \implicit\paren{\point, \param} = 0}$. Then, using the level set equation, the normal velocity at $\point \in \boundary\paren{\param}$ becomes \citep{stam2011velocity,osher2005level}
\begin{equation}\label{eqn:normal_velocity}
   \normalDispField\paren{\point, \param} = \frac{\paramderiv{\implicit}\paren{\point, \param}}{\norm{\nabla\implicit\paren{\point, \param}}}.
\end{equation}
\Cref{sec:geometric_examples} shows optimization of an implicit surface represented as a sum of radial basis functions (harmonic monopoles).

\setlength{\intextsep}{0pt} 
\setlength{\columnsep}{0.5em} 
\begin{wrapfigure}{r}{110pt}
\includegraphics{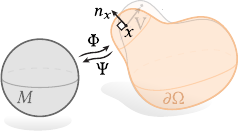}
\end{wrapfigure}

\subsubsection{Explicit surface}
\label{sec:ExplicitSurfaceRepresentation}
Here we imagine the boundary $\boundary\paren{\param}$ is parameterized over a \emph{reference surface} \(\reference\).  In particular, we view \(\boundary\paren{\param}\) as the image of a \emph{parameterization} $\embedding\paren{\cdot,\param}: \reference \to \R^3$; that is $\boundary\paren{\param} \coloneqq \curly{\embedding\paren{\pointref, \param} : \pointref \in \reference}$.  Concretely, one can think of \(\embedding\) as assigning coordinates to vertices of a mesh, or to control handles of a spline (\cref{sec:geometric_examples}).  We assume that \(\embedding\) is always a diffeomorphism, and hence has a well-defined inverse $\invEmbedding\paren{\cdot,\param}: \boundary\paren{\param} \to \reference$. In this case, the normal velocity at $\point \in \boundary\paren{\param}$ is
\begin{equation}
	\normalDispField\paren{\point, \param} = \normal_{\point} \cdot \paramderiv{\embedding}\paren{\invEmbedding\paren{\point,\param},\param}.
\end{equation}
Note that the normal velocity is independent of the specific choice of parameterization \citep{grinfeld2013introduction}.

\subsection{Screened Poisson equation}\label{sec:poisson}
The main PDE we consider is a \emph{screened Poisson} boundary value problem.  For simplicity we focus mainly on Dirichlet boundary conditions; \cref{sec:robin} treats Neumann and Robin boundary conditions.
\begin{myTitledBox}{\Forwardbvp{} boundary value problem}{primalBackground}{primalForeground}
	\begin{equation}
		\label{eqn:bvp}
		\begin{array}{rcll}
			\Delta \solution\paren{\point} - \screening \solution\paren{\point} &=& \source\paren{\point}	& \text{ in } \shape\paren{\param}, \\
			\solution\paren{\point} &=& \dirichlet\paren{\point, \param} & \text{ on } \boundary\paren{\param}. \\
		\end{array}
	\end{equation}
\end{myTitledBox}
\noindent Here $\Delta$ is the negative-semidefinite Laplace operator on $\R^3$, $\screening \in \R_{\ge 0}$ is a constant \emph{screening coefficient}, $\source : \domain\paren{\param} \to \R$ is the \emph{source term};
$\dirichlet\paren{\cdot, \param}: \boundary\paren{\param} \to \R$ is the \emph{boundary data}, which may also depend on parameters $\param$ (\cref{sec:bvp_deriv}), and $\solution: \domain\paren{\param} \to \R$ is the \emph{solution} of the BVP.
Although $\solution$ implicitly depends on the parameters $\param$, we typically omit this dependence to simplify notation.
We refer to \cref{eqn:bvp} as the \emph{\forwardbvp{} boundary value problem}, to distinguish it from the BVP developed in \cref{sec:bvp_deriv} to express derivatives. 

\subsection{Walk on spheres} \label{sec:wos}

\emph{Walk on spheres} (WoS) algorithms \citep{Muller:1956:WOS,Sawhney:2020:MCG} can be used to estimate solutions to PDEs such as \cref{eqn:bvp}.  The basic idea is to express the solution \(\solution\paren{\point_0}\) at a point \(\point_0 \in \domain\) as an integral over a ball around \(\point_0\), in terms of the unknown function \(\solution\).  Applying a single-sample Monte Carlo estimate to this integral then requires evaluating \(\solution\paren{\point_1}\), leading to a recursive estimation procedure (akin to Monte Carlo path tracing).  The random walk \(\point_0, \point_1, \ldots\) terminates when \(\point_k\) gets within a small distance \(\varepsilon > 0\) of the domain boundary \(\boundary\), where the solution is approximated using the value of \(\dirichlet\) at the closest boundary point.  Here we describe only the basic WoS estimator needed for our screened Poisson equation; see references in \cref{sec:walk-on-spheres-related} for further background.

\subsubsection{Notation}
\label{sec:walk-on-spheres-notation}

For any point $\point \in \domain\paren{\param}$, we let $\nearest{\point}$ be the closest point on the boundary $\boundary\paren{\param}$, and let $\distance{\point} \equiv \norm{\point - \nearest{\point}}$ be the corresponding shortest distance. We use $\ball\paren{\point, \mathsf{R}}$ for the ball with center $\point$ and radius $\mathsf{R}$, and $\maxball{\point} \equiv \ball(\point, \distance{\point})$ for the largest such ball contained in $\domain\paren{\param}$. We define the \emph{$\varepsilon$-shell} of $\boundary\paren{\param}$ as the set $\boundary^\varepsilon\paren{\param} \equiv \{\point \in \domain\paren{\param}: \distance{\point} \le \varepsilon\}$. Finally, we denote by $\green, \poisson: \R_{\geq 0} \to \R$ the Green's function and Poisson kernel (\resp) for the zero-Dirichlet screened Poisson equation on a ball \citep[Appendix A]{WoSt}; since these functions are rotationally symmetric, we parameterize them in terms of a positive radius.

\subsubsection{WoS estimator for Screened Poisson equations}

We can write the recursive WoS estimator for \cref{eqn:bvp} at point $\point_0$ as follows:
\begin{myTitledBox}{Walk on spheres estimator}{primalBackground}{primalForeground}
\vspace{-0.25em}
\begin{equation}
	\label{eqn:wos}
	\!\!\!\!\!\widehat{\solution}\paren{\point_k} \!\coloneqq\! \begin{cases}
		\dirichlet\paren{\nearest{\point_k}}, \!\!&\!\!\! \point_k \!\!\in\! \boundary^\varepsilon, \\[0.25em]
		\frac{\poisson(\distance{\point_k})}{\pdf^{\partial\ball}(\point_k)}\widehat{\solution}(\point_{k+1}) \!+\! \frac{\green(\norm{\pointaltalt_k - \point_k})}{\pdf^{\ball}(\point_k)}\source(\pointaltalt_k), \!\!&\!\!\! \text{otherwise}. \\
	\end{cases}
\end{equation}
\end{myTitledBox}
\noindent The next walk point $\point_{k+1}$ and source point $\pointaltalt_{k}$ are sampled from the sphere $\partial\maxball{\point_k}$ and ball $\maxball{\point_k}$ with uniform probability $\pdf^{\partial\ball}(\point_k) \equiv \nicefrac{1}{|\partial\maxball{\point_k}|}$ and $\pdf^{\ball}(\point_k) \equiv \nicefrac{1}{|\maxball{\point_k}|}$ (resp.).

\paragraph{Russian roulette} When $\screening > 0$, contributions at each walk step are attenuated by a multiplicative factor $\alpha \in [0, 1)$ (\emph{line 12}, \cref{alg:wos}). We apply Russian roulette \citep[Section 2.2.4]{Pharr:2023:PBR} proportionally to $\alpha$ to stochastically terminate walks before they reach  $\boundary^\varepsilon\paren{\param}$. Early termination improves the efficiency of the WoS estimator, without introducing additional bias. We summarize the resulting WoS procedure with Russian roulette in \cref{alg:wos}.

\subsection{Mixed Dirichlet-Robin boundary conditions}\label{sec:robin}
We can generalize our theory and algorithms to screened Poisson equations with mixed Dirichlet and Robin boundary conditions. Suppose we partition the domain boundary $\boundary\paren{\param}$ into a \emph{Robin boundary} $\boundaryNeumann$ and a \emph{Dirichlet boundary} $\boundaryDirichlet\paren{\param}$.  Our BVP becomes
\begin{equation}
	\label{eqn:mixed_bvp}
	\begin{array}{rcll}
		\Delta \solution\paren{\point} - \screening \solution\paren{\point} &=& \source\paren{\point}	& \text{ in } \shape\paren{\param}, \\
		\solution\paren{\point} &=& \dirichlet\paren{\point, \param} & \text{ on } \boundaryDirichlet\paren{\param}, \\
		\frac{\partial \solution}{\partial\normal}\paren{\point} - \robin\solution\paren{\point} &=& \neumann\paren{\point} & \text{ on } \boundaryNeumann. \\
	\end{array}
\end{equation}
Here $\nicefrac{\partial}{\partial \normal}$ is the normal derivative; $\robin \in \R_{\ge 0}$ is a constant \emph{Robin coefficient}; $\neumann: \boundaryNeumann \to \R$ is the \emph{Robin data}; for simplicity, we assume $\robin$, $\neumann$, and $\boundaryNeumann$ do not depend on $\param$.  The solution to \cref{eqn:mixed_bvp} can be estimated via the \emph{walk on stars (WoSt)} method \citep{WoStRobin}, which is very similar to WoS but considers star-shaped domains rather than balls (see \citet[Algorithm 1]{WoStRobin} for details).  We give a \diffbvp{} version of this estimator in \cref{sec:diffwost}.

\begin{algorithm}[t]
	\caption{\Proc{WalkOnSpheres}$(\point,\ \varepsilon)$}
	\label{alg:wos}
	\begin{algorithmic}[1]
	\algblockdefx[Name]{Class}{EndClass}
		[1][Unknown]{\textbf{class} #1}
		{}
	\algtext*{EndClass}
	\algblockdefx[Name]{FORDO}{ENDFORDO}
		[1][Unknown]{\textbf{for} #1 \textbf{do}}
		{}
	\algtext*{ENDFORDO}
	\algblockdefx[Name]{IF}{ENDIF}
		[1][Unknown]{\textbf{if} #1 \textbf{then}}
		{}
	\algtext*{ENDIF}
	\algblockdefx[Name]{IFTHEN}{ENDIFTHEN}
		[2][Unknown]{\textbf{if} #1 \textbf{then} #2}
		{}
	\algtext*{ENDIFTHEN}
	\algblockdefx[Name]{RETURN}{ENDRETURN}
		[1][Unknown]{\textbf{return} #1}
		{}
	\algtext*{ENDRETURN}
	\algblockdefx[Name]{COMMENT}{ENDCOMMENT}
		[1][Unknown]{\textcolor{primalComment}{\(\triangleright\) #1}}
		{}
	\algtext*{ENDCOMMENT}

	\Require A point $\point$, and a termination parameter $\varepsilon$.
	\Ensure A single-sample estimate $\widehat{\solution}(\point)$ for \cref{eqn:bvp}.

		\COMMENT[Compute distance to and closest point on $\boundary$]\ENDCOMMENT
		\State $\distanceEmpty,\ \nearest{\point} \gets \Proc{DistanceToBoundary}(\point)$

		\COMMENT[Return boundary value $\dirichlet$ at $\nearest{\point}$ if $\point \in \boundary^{\varepsilon}$]\ENDCOMMENT
		\IFTHEN[$\distanceEmpty < \varepsilon$]{\textbf{return} $\dirichlet(\nearest{\point})$}
		\ENDIFTHEN

		\COMMENT[Uniformly sample a point $\pointball$ inside the unit ball]\ENDCOMMENT
		\State $\pointball \gets \Proc{SampleUnitBall()}$

		\COMMENT[Compute point for source contribution]\ENDCOMMENT
		\State $\pointaltalt \gets \point + \distanceEmpty \cdot \pointball$

		\COMMENT[Compute source contribution]\ENDCOMMENT
		\State $\widehat{S} \gets \green(\norm{\pointaltalt-\point}) / \pdf^{\ball}(\point) \cdot \source(\pointaltalt)$

		\COMMENT[Compute recursive contribution scale]\ENDCOMMENT
		\State $\alpha \gets \poisson(\distanceEmpty) / \pdf^{\partial\ball}(\point)$

		\COMMENT[Attempt to terminate walk using Russian roulette]\ENDCOMMENT
		\IFTHEN[$\alpha < \Proc{SampleUniform(0, 1)}$]{\textbf{return} $\widehat{S}$}
		\ENDIFTHEN

		\COMMENT[Uniformly sample a point $\pointsphere$ on the unit sphere]\ENDCOMMENT
		\State $\pointsphere \gets \Proc{SampleUnitSphere()}$

		\COMMENT[Compute next walk point]\ENDCOMMENT
		\State $\point \gets \point + \distanceEmpty \cdot \pointsphere$

		\COMMENT[Repeat procedure from next walk point]\ENDCOMMENT
		\RETURN[$\Proc{WalkOnSpheres}(\point, \varepsilon) + \widehat{S}$]
		\ENDRETURN
	\end{algorithmic}
\end{algorithm}

\section{PDE-constrained shape optimization}\label{sec:optimization}
This section defines our central optimization problem (\cref{eqn:functional}).  We use classical results from shape optimization \citep{henrot2018shape} to express the derivatives in this problem as another BVP (\cref{sec:bvp_deriv}), which we can then estimate via WoS (\cref{sec:diffwos}).

\paragraph{Shape functional} We consider a domain $\domain\paren{\param}$ controlled by parameters $\param$ as in \cref{sec:continuum}, and the solution $\solution\paren{\cdot, \param}$ to the \forwardbvp{} BVP \labelcref{eqn:bvp} on this domain. We want to determine values for $\param$ that locally minimize a \emph{shape functional}
\begin{equation}
   \label{eqn:functional}
   \shapeObjective\paren{\param} \equiv \int_{\domain\paren{\param}} \Mask\paren{\point}\solnLoss\paren{\solution\paren{\point,\param}} \ud \point.
\end{equation}
Here $\solnLoss: \R \to \R$ is a differentiable \emph{loss function}---for example, the squared loss $\solnLoss\paren{\solution\paren{\point,\param}} \coloneqq \nicefrac{1}{2}\norm{\solution\paren{\point,\param} - \solution_{\mathrm{ref}}\paren{\point}}^2$ relative to a reference solution $\solution_{\mathrm{ref}}$.  The function $\Mask: \R^3 \to \curly{0,1}$ is a \emph{binary mask} that we use to localize the functional to a subdomain of $\domain\paren{\param}$, such as a region of interest where a reference is available (see \cref{sec:geometric_examples}). \cref{app:shape_functional} considers a more general functional that includes a boundary term, used in some of our examples (\cref{sec:geometric_examples}).

\paragraph{Differentiating shape functionals} We minimize \(\shapeObjective\paren{\param}\) using \emph{stochastic} gradient optimization, which requires stochastic estimates of the derivative of the shape functional $\shapeObjective\paren{\param}$ with respect to the parameters $\param$. From the Reynolds transport theorem, this derivative equals \citep[p. 239]{henrot2018shape}:%
\footnote{Throughout, we distinguish between \emph{total derivatives} $\nicefrac{\dd}{\dd\param}$ and \emph{partial derivatives} $\nicefrac{\partial}{\partial\param}$.}
\begin{align}\label{eqn:functional_deriv}
	\frac{\dd \shapeObjective}{\dd \param}\paren{\param} &= \int_{\domain\paren{\param}} \Mask\paren{\point}\ \solnDeriv\paren{\point, \param}\solnLossDeriv(\solution\paren{\point, \param}) \ud \point \nonumber \\
	&+ \int_{\boundary\paren{\param}} \Mask\paren{\point}\normalDispField\paren{\pointalt, \param}\solnLoss(\solution\paren{\pointalt, \param}) \ud \pointalt\paren{\pointalt},
\end{align}
where $\solnLossDeriv$ is the derivative of the scalar loss $\solnLoss$. \Cref{eqn:functional_deriv} is valid as long as the integrand of \(\shapeObjective\) does not have any $\param$-dependent discontinuities---this assumption holds for all $\param$-independent masking functions $\Mask$ and smooth loss functions $\solnLoss$, as $\solution$ is itself smooth.  

Note that the boundary of \(\Omega\) is always closed; in \cref{sec:discussion} we discuss how this formulation might be extended to open boundaries.

\subsection{The \diffbvp{} boundary value problem}
\label{sec:bvp_deriv}

Evaluating \cref{eqn:functional_deriv} requires computing the derivative $\solnDeriv$ of the the solution $\solution$ with respect to parameters $\param$. In this section, we express $\solnDeriv$ as the solution to a different BVP, which can in turn be estimated via a modified WoS algorithm (\cref{sec:diffwos}).

The \forwardbvp{} BVP \labelcref{eqn:bvp} implicitly defines $\solution$ as a function of $\param$. As explained by \citet[Section 5.3]{henrot2018shape}, we can thus use \emph{implicit differentiation} to obtain the derivatives $\solnDeriv$ of this relationship. To this end, we first differentiate both sides of the screened Poisson equation with respect to \(\param\). Since the Laplacian \(\Delta\), screening coefficient \(\screening\), and source term \(\source\) are independent of $\param$, we get
\begin{equation}\label{eqn:poisson_deriv}
	\Delta \solnDeriv\paren{\point} - \screening \solnDeriv\paren{\point} = 0	\quad \text{ in } \shape.
\end{equation}
Implicit differentiation of the boundary condition depends on how we represent the boundary data $\dirichlet$. We consider two cases.

\paragraph{Restricted boundary data} We first consider the case where $\dirichlet$ is the restriction of a scalar field $\dirichletField\paren{\cdot, \param}: \R^3 \to \R$ to the domain boundary, \ie, $\dirichlet\paren{\point} = \dirichletField\paren{\point, \param}, \forall \point \in \boundary\paren{\param}$. Differentiating the boundary condition in \cref{eqn:bvp} with respect to \(\param\) then yields
\begin{equation}\label{eqn:boundary_deriv_restricted}
	\solnDeriv\paren{\point} = \dirichletFieldDeriv\paren{\point} + \paren{\frac{\partial \dirichlet}{\partial \normal}\paren{\point} - \frac{\partial \solution}{\partial \normal}\paren{\point}}\normalDispField\paren{\point, \param} \quad \text{ on } \boundary\paren{\param},
\end{equation}
where \(\nicefrac{\partial \dirichlet}{\partial \normal} = \normal_{\point} \cdot \nabla\dirichletField\) (\cref{app:proof}).

\paragraph{Mapped boundary data} When we use an explicit representation for $\boundary\paren{\param}$ (\cref{sec:ExplicitSurfaceRepresentation}), we can alternatively define boundary values \(\dirichletReference(\cdot,\param): \reference \to \R\) on the reference surface \(\reference\) (possibly depending on \(\param\)).  For instance, if \(\reference\) is a mesh, then \(\dirichletReference\) could be defined via a texture map.  These values are then pushed forward to \(\boundary\paren{\param}\) to obtain the current boundary data, \ie, \(\dirichlet\paren{\point} = \dirichletReference\paren{\invEmbedding\paren{\point,\param},\param}\).  In \cref{app:proof} we show that differentiating the boundary condition in \cref{eqn:bvp} yields
\begin{align}\label{eqn:boundary_deriv_mapped}
	\solnDeriv\paren{\point} &= \dirichletReferenceDeriv\paren{\invEmbedding\paren{\point}} + \nabla \dirichletReference\paren{\invEmbedding\paren{\point}} \cdot
	\paramderiv{\invEmbedding}\paren{\point} \nonumber \\
	&+ \paren{\frac{\partial \dirichlet}{\partial \normal}\paren{\point} - \frac{\partial \solution}{\partial \normal}\paren{\point}}\normalDispField\paren{\point, \param} \quad \text{ on } \boundary\paren{\param},
\end{align}
where $\nabla\dirichletReference$ is the spatial gradient of $\dirichletReference$ on the reference surface, and  \(\nicefrac{\partial \dirichlet}{\partial \normal} = \nabla\dirichletReference \cdot (J_{\invEmbedding} \normal_{\point})\), where $J_{\invEmbedding}$ is the Jacobian of \(\invEmbedding\).

\paragraph{Summary}
Combining \cref{eqn:poisson_deriv,eqn:boundary_deriv_restricted,eqn:boundary_deriv_mapped}, we can express the derivative $\solnDeriv$ of the BVP solution $\solution$ with respect to the parameters $\param$ as the solution to the following BVP:
\begin{myTitledBox}{\Diffbvp{} boundary value problem}{differentialBackground}{differentialForeground}
\begin{equation}
	\label{eqn:bvp_deriv}
	\begin{array}{rcll}
		\Delta \solnDeriv\paren{\point} - \screening \solnDeriv\paren{\point} \!\!\!\!\!&=&\!\!\!\!\! 0	\!\!&\!\! \text{ in } \shape\paren{\param}, \\
		\solnDeriv\paren{\point} \!\!\!\!\!&=&\!\!\!\!\! \dirichletDeriv\paren{\point, \param} \!-\!\normalDispField\paren{\point, \param} \frac{\partial \solution}{\partial \normal}\paren{\point} \!\!&\!\! \text{ on } \boundary\paren{\param}. \\
	\end{array}
\end{equation}
\end{myTitledBox}
\noindent In \cref{eqn:bvp_deriv}, the function $\dirichletDeriv$ can be inferred from the right-hand side of either \cref{eqn:boundary_deriv_restricted} or \cref{eqn:boundary_deriv_mapped}. We call \cref{eqn:bvp_deriv} the \emph{\diffbvp{} boundary value problem}. Comparing the \diffbvp{} BVP with the \forwardbvp{} BVP (\cref{eqn:bvp}), we observe the following:
\begin{enumerate}[leftmargin=*,nosep,label={{O\arabic*}.},ref={{O\arabic*}}]
	\item\label{enu:type} Both BVPs solve the screened Poisson equation with Dirichlet boundary conditions in the same domain, differing only in their source term and boundary data.
	\item\label{enu:coupling} The two BVPs are \emph{nested}, as the boundary data for \(\solnDeriv\) in the \diffbvp{} BVP depends on the solution \(\solution\) to the \forwardbvp{} BVP.
\end{enumerate}
These observations will facilitate the derivation of Monte Carlo estimators for the \diffbvp{} BVP in \cref{sec:diffwos}.

\section{Monte Carlo derivative estimation}\label{sec:derivatives}

We now develop Monte Carlo algorithms for estimating the shape-functional derivative $\nicefrac{\dd \shapeObjective}{\dd \param}$ in \cref{eqn:functional_deriv}. We do this in two parts: First, in \cref{sec:diffwos}, we introduce a \diffbvp{} walk on spheres algorithm for computing point estimates of $\paramderiv{\solution}\paren{\cdot,\param}$. Then, in \cref{sec:backward}, we explain how to use these estimates to form Monte Carlo estimates of the domain and boundary integrals in \cref{eqn:functional_deriv}.

The differentiable rendering literature often refers to estimates of $\paramderiv{\solution}$---and in particular their evaluation on a dense sampling of the domain $\domain\paren{\param}$---as \emph{forward-mode derivatives} \citep{nimier2020radiative,vicini2021path,zhang2023projective}. Our \diffbvp{} walk on spheres algorithm in \cref{sec:diffwos} allows computing such derivatives, which we visualize in \cref{fig:forward_derivatives}, \ref{fig:finite_difference}, and \ref{fig:ablation} to evaluate variance and bias. However, as we explain in \cref{sec:backward}, in optimization settings we can use our algorithm to directly compute so-called \emph{reverse-mode (backward) derivatives} $\nicefrac{\dd \shapeObjective}{\dd \param}$ with respect to all parameters $\param$, without storing intermediate forward-mode derivatives. Doing so is analogous to the radiative backpropagation procedure by \citet{nimier2020radiative} and \citet{vicini2021path}.

\subsection{\Diffbvp{} walk on spheres}\label{sec:diffwos}

Observations~\labelcref{enu:type,enu:coupling} provide guidance for how to develop a Monte Carlo algorithm for estimating $\paramderiv{\solution}$. Observation \labelcref{enu:type} shows that the WoS estimator we presented in \cref{sec:wos}, for the solution $\solution$ to the \forwardbvp{} BVP \labelcref{eqn:bvp}, can also estimate the solution $\solnDeriv$ to the \diffbvp{} BVP \labelcref{eqn:bvp_deriv}, with a caveat: Observation \labelcref{enu:coupling} shows we must \emph{nest} a WoS estimator for the \forwardbvp{} BVP within one for the \diffbvp{} BVP---that is, allow the latter to use the outputs of the former.

The fact that (the normal derivative of) $\solution$ only appears in the Dirichlet boundary data of the \diffbvp{} BVP \cref{eqn:bvp_deriv} makes this nesting straightforward: Once the WoS estimator for $\solnDeriv$ reaches some point inside $\boundary^\varepsilon\paren{\param}$, to compute the Dirichlet boundary data needed for termination, it invokes the WoS estimator for $\nicefrac{\partial \solution}{\partial n}$, as we describe below. This nesting is equivalent to a process that starts a random walk for $\solnDeriv$ and, at the point where the walk terminates, starts another random walk for $\nicefrac{\partial \solution}{\partial n}$ (\cref{fig:coupling}).

\paragraph{Estimating normal derivatives} Complicating slightly this nesting procedure is the fact that the boundary condition of the \diffbvp{} BVP \labelcref{eqn:bvp_deriv} uses not the solution $\solution$, but its normal derivative $\nicefrac{\partial \solution}{\partial \normal}$ at a boundary point $\point \in \boundary\paren{\param}$. A modified WoS algorithm by \citet[Section 3]{Sawhney:2020:MCG} can estimate spatial gradients; however, as \citet[Section 3.2]{BVC} explain, it cannot do so at boundary points. Instead, \citeauthor{BVC} approximate the normal derivative at $\point$ at an \emph{offset point} $\point - \offset \cdot \normal_{\point}$ with $\offset > \varepsilon$, which they then estimate with the modified WoS algorithm \citep[Equation 13]{Sawhney:2020:MCG}. The result is the overall estimator
\begin{equation}
	\label{eqn:dudn_mc_estimate}
	\widehat{\frac{\partial \solution}{\partial \normal}}\paren{\point} \stackrel{\text{\citep{BVC}}}{\coloneqq} \widehat{\frac{\partial \solution}{\partial \normal}}\paren{\point - \offset\cdot n_x}
	= \frac{3}{\offset}\paren{\normal_\pointalt \cdot \normal_\point} \cdot \widehat{\solution}\paren{\pointalt},
\end{equation}
where the point $\pointalt$ is sampled uniformly on the sphere $\partial\maxball{\point - \offset\cdot \normal_x}$.

We instead use backward differences to estimate the normal derivative through an estimate of the \emph{solution} at the offset point as
\begin{equation}
	\label{eqn:dudn_fd_estimate}
	\widehat{\frac{\partial \solution}{\partial \normal}}\paren{\point} \coloneqq
	\frac{\dirichlet\paren{\point} - \widehat{\solution}\paren{\point-\offset\cdot\normal_\point}}{\offset}.
\end{equation}
In \cref{eqn:dudn_mc_estimate,eqn:dudn_fd_estimate}, $\widehat{\solution}$ uses the WoS estimator of \cref{eqn:wos}. We found empirically (\cref{fig:ablation}) that our approach results in similar bias but less variance than the approach of \citet{BVC}. We suspect that the backward-differences estimator reduces variance by avoiding a cosine term over the first ball of the recursive walk. A more comprehensive comparison of derivative estimators near the boundary, including off-centered gradient estimators~\citep[Equation 26]{Yu:2024:DiffPoisson}, would be a useful future experiment.

\begin{figure}[t]
	\centering
	\includegraphics[width=\columnwidth]{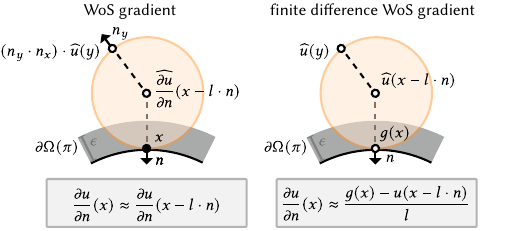}
	\caption{\figloc{Left:} The WoS estimator for the normal derivative by \citet{Sawhney:2020:MCG} in \cref{eqn:dudn_mc_estimate} is undefined for points on the boundary, $\point\in\boundary\paren{\param}$, so \citet{BVC} evaluate it at an offset point $\point - \offset\cdot\normal$. \figloc{Right:} We instead use the backward-difference approximation of the normal derivative in \cref{eqn:dudn_fd_estimate} at the offset point.}
	\label{fig:boundary_gradient}
\end{figure}

\paragraph{Final estimator} Starting at the location $\point_0 = \point$ where we want to estimate $\solnDeriv$, we can write our estimator recursively as follows:
\begin{myTitledBox}{\Diffbvp{} walk on spheres estimator}{differentialBackground}{differentialForeground}
\vspace{-0.25em}
\begin{equation}
	\label{eqn:wos_deriv}
	\!\!\!\!\!\widehat{\solnDeriv}\paren{\point_k}\! \coloneqq\! \begin{cases}
		\dirichletDeriv(\nearest{\point_k}) \!-\! \frac{\normalDispField(\nearest{\point_k})}{\offset}(\dirichlet(\nearest{\point_k}) \!-\!\widehat{\solution}(\nearest{\point_k} - \offset \normal_{\nearest{\point_k}})), \!\!&\!\!\! \point_k \!\!\in\! \boundary^\varepsilon, \\[0.25em]
		\frac{\poisson(\distance{\point_k})}{\pdf^{\partial\ball}(\point_k)}\widehat{\solnDeriv}(\point_{k+1}), \!\!&\!\!\! \text{otherwise}. \\
	\end{cases}
\end{equation}
\end{myTitledBox}
\noindent In \cref{eqn:wos_deriv}, $\widehat{\solution}$ is the WoS estimator of \cref{eqn:wos}. As in WoS, the next walk point $\point_{k+1}$ is sampled on the sphere $\partial\maxball{\point_k}$ with uniform probability $\pdf^{\partial\ball}\paren{\point_k}$. We term \cref{eqn:wos_deriv} the \emph{\diffbvp{} walk on spheres} estimator. As in WoS, we can also combine this estimator with Russian roulette for improved efficiency; we summarize the resulting procedure in \cref{alg:wos_deriv}.

We conclude with two observations about the estimator in \cref{eqn:wos_deriv}:
\begin{enumerate*}
	\item Whereas at any point $\point$ the solution $\solution\paren{\point}$ is a scalar, the derivative $\solnDeriv\paren{\point}$ is an $N$-dimensional vector, where $N$ is the number of parameters $\param$. \Cref{alg:wos_deriv} estimates the derivatives with respect to all parameters $\param$ with a \emph{single walk}.
	\item Along with a single-sample estimate of $\solnDeriv\paren{\point}$, \cref{alg:wos_deriv} can provide a single-sample estimate of $\solution\paren{\point}$ by returning the Dirichlet boundary condition at the terminal point $\dirichlet(\nearest{\point})$ in \cref{line:wos_deriv_return}---effectively, using the same walk for both $\solnDeriv\paren{\point}$ and $\solution\paren{\point}$. We do not make this modification explicit in \cref{alg:wos_deriv} to keep the basic algorithm simple, but we use it for estimation of reverse-mode derivatives in \cref{sec:backward}.
\end{enumerate*}

\begin{algorithm}[t]
	\caption{\Proc{DiffWalkOnSpheres}$(\point,\ \varepsilon,\ \offset)$}
	\label{alg:wos_deriv}
	\begin{algorithmic}[1]
	\algblockdefx[Name]{Class}{EndClass}
		[1][Unknown]{\textbf{class} #1}
		{}
	\algtext*{EndClass}
	\algblockdefx[Name]{FORDO}{ENDFORDO}
		[1][Unknown]{\textbf{for} #1 \textbf{do}}
		{}
	\algtext*{ENDFORDO}
	\algblockdefx[Name]{IF}{ENDIF}
		[1][Unknown]{\textbf{if} #1 \textbf{then}}
		{}
	\algtext*{ENDIF}
	\algblockdefx[Name]{IFTHEN}{ENDIFTHEN}
		[2][Unknown]{\textbf{if} #1 \textbf{then} #2}
		{}
	\algtext*{ENDIFTHEN}
	\algblockdefx[Name]{RETURN}{ENDRETURN}
		[1][Unknown]{\textbf{return} #1}
		{}
	\algtext*{ENDRETURN}
	\algblockdefx[Name]{COMMENT}{ENDCOMMENT}
		[1][Unknown]{\textcolor{differentialComment}{\(\triangleright\) #1}}
		{}
	\algtext*{ENDCOMMENT}
	\algblockdefx[Name]{COMMENTPRIMAL}{ENDCOMMENTPRIMAL}
		[1][Unknown]{\textcolor{primalComment}{\(\triangleright\) #1}}
		{}
	\algtext*{ENDCOMMENTPRIMAL}

	\Require A point $\point$, a termination parameter $\varepsilon$, and an offset $\offset$.
	\Ensure A single-sample estimate $\widehat{\solnDeriv}(\point)$ for \cref{eqn:bvp_deriv}.

		\COMMENT[Compute distance to and closest point on $\boundary$]\ENDCOMMENT
		\State $\distanceEmpty,\ \nearest{\point} \gets \Proc{DistanceToBoundary}(\point)$

		\COMMENT[Compute and return boundary value at $\nearest{\point}$ if $\point \in \boundary^{\varepsilon}$]\ENDCOMMENT
		\IF[$\distanceEmpty < \varepsilon$]
			\COMMENTPRIMAL[Compute outward normal at $\nearest{\point}$]\ENDCOMMENTPRIMAL
			\State $\normal \gets \Proc{BoundaryNormal}(\nearest{\point})$

			\COMMENTPRIMAL[Compute offset point]\ENDCOMMENTPRIMAL
			\State $\pointoff \gets \nearest{\point} - \offset \normal$

			\COMMENTPRIMAL[Estimate \forwardbvp{} BVP solution at offset point]\ENDCOMMENTPRIMAL
			\State $\widehat{\solution} \gets \Proc{WalkOnSpheres}(\pointoff, \varepsilon)$

			\COMMENTPRIMAL[Return boundary value at $\nearest{\point}$]\ENDCOMMENTPRIMAL
			\textbf{return} $\dirichletDeriv(\nearest{\point}) - \normalDispField(\nearest{\point})(\dirichlet(\nearest{\point}) - \widehat{\solution}) / \offset$\label{line:wos_deriv_return}
		\ENDIF

		\COMMENT[Compute recursive contribution scale]\ENDCOMMENT
		\State $\alpha \gets \poisson(\distanceEmpty) / \pdf^{\partial\ball}(\point)$

		\COMMENT[Attempt to terminate walk using Russian roulette]\ENDCOMMENT
		\IFTHEN[$\alpha < \Proc{SampleUniform(0, 1)}$]{\textbf{return} $0$}
		\ENDIFTHEN

		\COMMENT[Uniformly sample a point $\pointsphere$ on the unit sphere]\ENDCOMMENT
		\State $\pointsphere \gets \Proc{SampleUnitSphere()}$

		\COMMENT[Compute next walk point]\ENDCOMMENT
		\State $\point \gets \point + \distanceEmpty \cdot \pointsphere$

		\COMMENT[Repeat procedure from next walk point]\ENDCOMMENT
		\RETURN[$\Proc{DiffWalkOnSpheres}(\point, \varepsilon, \offset)$]
		\ENDRETURN
	\end{algorithmic}
\end{algorithm}

\subsection{\Diffbvp{} walk on stars}\label{sec:diffwost}

We now return to the BVP \labelcref{eqn:mixed_bvp} with mixed Dirichlet-Robin boundary conditions. We can derive a Monte Carlo estimator for the derivative $\solnDeriv$ of its solution $\solution$ with respect to parameters $\param$, exactly analogously to how we did so for the \forwardbvp{} BVP \labelcref{eqn:bvp} in \cref{sec:bvp_deriv,sec:diffwos}. In particular, $\solnDeriv$ is the solution to the BVP \citep{henrot2018shape}
\begin{equation}
	\label{eqn:mixed_bvp_deriv}
	\begin{array}{rcll}
		\Delta \solnDeriv\paren{\point} - \screening \solnDeriv\paren{\point} \!\!\!\!\!&=&\!\!\!\!\! 0	\!\!&\!\! \text{ in } \shape\paren{\param}, \\
		\solnDeriv\paren{\point} \!\!\!\!\!&=&\!\!\!\!\! \dirichletDeriv\paren{\point, \param} \!-\!\normalDispField\paren{\point, \param} \frac{\partial \solution}{\partial \normal}\paren{\point} \!\!&\!\! \text{ on } \boundaryDirichlet\paren{\param}, \\
		\frac{\partial \solnDeriv}{\partial\normal}\paren{\point} - \robin\solnDeriv\paren{\point} \!\!\!\!\!&=&\!\!\!\!\! 0 \!\!&\!\! \text{ on } \boundaryNeumann. \\
	\end{array}
\end{equation}
The observations \labelcref{enu:type} and \labelcref{enu:coupling} we made for the Dirichlet-only \forwardbvp{} and \diffbvp{} BVPs (\cref{eqn:bvp,eqn:bvp_deriv}) also hold for their mixed Dirichlet-Robin counterparts (\cref{eqn:mixed_bvp,eqn:mixed_bvp_deriv}). Importantly, in the latter case nesting happens through \emph{only} the Dirichlet boundary condition. Thus, we can estimate $\solnDeriv$ using a \diffbvp{} WoSt algorithm that relates to the original WoSt exactly analogously to the relationship between \diffbvp{} and original WoS (\cref{alg:wos_deriv,alg:wos}): Namely, the \diffbvp{} WoSt estimator performs a random walk to estimate $\solnDeriv$, until it reaches a point in the $\varepsilon$-shell around $\boundaryDirichlet$; there, it invokes the WoSt estimator, which performs another random walk to estimate $\nicefrac{\partial \solution}{\partial \normal}$. In \cref{sec:thermal_design}, we use this \diffbvp{} WoSt algorithm in a thermal design setting.

We emphasize that the above approach applies only to the case where the Robin boundary condition in \cref{eqn:mixed_bvp} is independent of the parameters $\param$ we optimize. We leave parameterizing and differentiating the Robin boundary condition to future work, but discuss challenges with such an extension in \cref{sec:discussion}.

\subsection{Computing reverse-mode derivatives}\label{sec:backward}
\begin{figure}[t]
	\centering
	\includegraphics[width=\columnwidth]{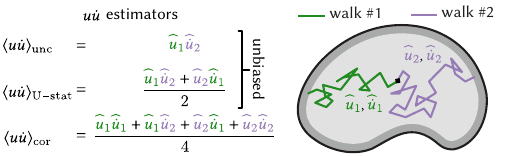}
	\caption{A Monte Carlo estimate of the product $\langle \solution\solnDeriv \rangle$ requires estimating both the solution $\solution$ and derivative $\solnDeriv$. A single sampled walk simultaneously provides estimates of $\solution$ and $\solnDeriv$, but the product estimates $\solution \solnDeriv$ are correlated and introduce bias. Rather than resort to uncorrelated estimates, which uses a single estimate from each walk, U-statistics shares complementary estimates across walks, which reduces variance without introducing bias.}
	\label{fig:product_estimation}
\end{figure}

We approximate the domain and boundary integrals in \cref{eqn:functional_deriv} using Monte Carlo integration, by uniformly sampling points in the domain $\domain\paren{\param}$ (for the first integral) or boundary $\boundary\paren{\param}$ (for the second integral), and evaluating the corresponding integrands at those locations. Evaluating the integrand at different sample points $\point \in \domain\paren{\param}$ requires computing the loss function derivative $\solnLossDeriv$, which in turn requires computing values for both $\solution$ and $\solnDeriv$. The option to share walks between the \forwardbvp{} and \diffbvp{} WoS estimators of these values creates opportunities for optimization. To elaborate, in the rest of this section, we specialize to the squared loss function $\solnLoss\paren{\solution\paren{\point}} \coloneqq \nicefrac{1}{2}\norm{\solution\paren{\point} - \solution_{\mathrm{ref}}\paren{\point}}^2$. Then
\begin{equation}
	\label{eqn:squared_loss_deriv}
	\frac{\dd\solnLoss}{\dd\param}\paren{\solution\paren{\point}} = \solution\paren{\point}\solnDeriv\paren{\point} - \solution_{\mathrm{ref}}\paren{\point}\solnDeriv\paren{\point}.
\end{equation}
We discuss two baseline approaches to estimate \cref{eqn:squared_loss_deriv}, and in particular the product $\solution\solnDeriv$. Then we propose our own improved approach. For this discussion, we assume we have performed $2M$ walks of \cref{alg:wos_deriv}, augmented as we described above so that each walk estimates both $\solution$ and $\solnDeriv$, with corresponding outputs $\curly{\uest_m, \uestdot_m}_{m=1}^{2M}$. We visualize all approaches in \cref{fig:product_estimation} for the case $M = 1$.

\paragraph{Uncorrelated product estimator} The first baseline approach uses disjoint sets of walks---equivalently, separate calls to \cref{alg:wos} and \cref{alg:wos_deriv}---to compute \emph{independent} estimates $\uest$ and $\uestdot$, then computes their product. We can express this approach as
\begin{equation}\label{eqn:uncorrelated_product}
	\uest \coloneqq \frac{1}{M} \sum_{m=1}^{M} \uest_m, \quad \uestdot \coloneqq \frac{1}{M} \sum_{m=M+1}^{2M} \uestdot_m, \quad \Estimator{\solution\solnDeriv}_{\text{unc}} \coloneqq \uest \cdot \uestdot.
\end{equation}
This approach is unbiased:%
\footnote{More precisely, it does not introduce additional bias beyond that from the \forwardbvp{} and \diffbvp{} WoS estimators (due to the $\varepsilon$-shell and normal-derivative approximations).}
from independence, $\E{\Estimator{\solution\solnDeriv}_{\text{unc}}} = \E{\uest} \cdot \E{\uestdot}$. However, it is sample-inefficient: it does not leverage the fact that each walk produces estimates for both $\solution$ and $\solnDeriv$, resulting in variance that scales as $\bigO\paren{\nicefrac{1}{M^2}}$.

\paragraph{Correlated product estimator} The second baseline approach uses all walks to compute \emph{correlated} estimates $\uest$ and $\uestdot$, then again computes their product. We can express this approach as
\begin{equation}\label{eqn:correlated_product}
	\uest \coloneqq \frac{1}{2M} \sum_{m=1}^{2M} \uest_m, \quad \uestdot \coloneqq \frac{1}{2M} \sum_{m=1}^{2M} \uestdot_m, \quad \Estimator{\solution\solnDeriv}_{\text{cor}} \coloneqq \uest \cdot \uestdot.
\end{equation}
Conversely to the uncorrelated approach, this approach is sample-efficient: it uses outputs for both $\solution$ and $\solnDeriv$ from all walks, and thus reduces variance at a faster rate $\bigO\paren{\nicefrac{1}{4 M^2}}$. However, it is biased: because of correlation, $\E{\Estimator{\solution\solnDeriv}_{\text{cor}}} \neq \E{\uest} \cdot \E{\uestdot}$.

\paragraph{Our approach: U-statistic product estimator} To combine unbiasedness and sample efficiency, we introduce an approach that uses U-statistics---a methodology from statistics \citep{lee1990ustatistics} for combining estimators input to symmetric functions (\eg, product) without introducing bias. Within computer graphics, \citet{kettunen2021unbiased} used U-statistics to evaluate power-series transmittance estimators.

The U-statistic product estimator uses all pairwise combinations of estimates $\uest_m$ and $\uestdot_{m'}$ such that $m \neq m'$
\begin{equation}
	\label{eqn:ustatistics_unrolled}
	\Estimator{\solution\solnDeriv}_{\text{U-stat}} \coloneqq \frac{1}{2M(2M-1)}\sum_{\vphantom{m'}m=1}^{2M} \sum_{\substack{m'=1\\m'\neq m}}^{2M} \uest_m\uestdot_{m'}.
\end{equation}
Compared to the correlated estimator $\Estimator{\solution\solnDeriv}_{\text{corr}}$ in \cref{eqn:correlated_product}, we observe that the U-statistic summation:
\begin{enumerate*}
	\item excludes only the $2M$ correlated product terms $\uest_m\uestdot_{m}$ to achieve unbiasedness, \ie, $\E{\Estimator{\solution\solnDeriv}_{\text{U-stat}}} = \E{\uest} \cdot \E{\uestdot}$;
	\item includes all other $2M(2M-1)$ product terms to achieve sample efficiency, reducing variance at a rate $\bigO\paren{\nicefrac{1}{4M^2}}$.
\end{enumerate*}
We can write \cref{eqn:ustatistics_unrolled} equivalently as
\begin{equation}
	\label{eqn:ustatistics}
	\Estimator{\solution\solnDeriv}_{\mathrm{U-stat}} = \frac{1}{2M (2M-1)}\sum_{m=1}^{2M} \uestdot_m \paren{S - \uest_m},\quad S\equiv\sum_{m=1}^{2M} \uest_m.
\end{equation}
\Cref{eqn:ustatistics} provides a way to compute the U-statistic estimator in $\bigO\paren{2M}$ time, resulting in negligible computational overhead compared to the correlated and uncorrelated estimator. However, this estimator introduces a memory overhead, due to the need to store $\uest_m$ and $\uestdot_m$ estimates as we sample walks. In our implementation, we strike a balance between sample efficiency and memory efficiency by applying U-statistics in batches of $B < 2M$ walks, computing \cref{eqn:ustatistics} after every $\nicefrac{2M}{B}$ walks. In \cref{sec:ablation}, we evaluate experimentally the performance of the U-statistic estimator compared to the uncorrelated and correlated ones, confirming it achieves sample efficiency while retaining unbiasedness.

\paragraph{Other loss functions} Our discussion assumed the squared loss function, but similar considerations apply to any other differentiable loss function $\solnLoss$: namely, we can apply the uncorrelated, correlated, and U-statistic estimators to the product $\nicefrac{\dd \solnLoss}{\dd \param} = \solnDeriv \solnLossDeriv\paren{\solution}$, instead of $\solnDeriv\solution$. In this case, none of the estimators will be unbiased, because in general $\E{\solnLossDeriv\paren{\solution}} \neq \solnLossDeriv\paren{\E{\solution}}$. However, we expect that the relative performance of the three estimators in terms of bias and sample efficiency will remain the same as in the case of the squared loss.

Our discussion on estimation of products $\solnDeriv \solnLossDeriv\paren{\solution}$ for reverse-mode differentiation finds exact parallels in differentiable rendering, where such products appear when differentiating inverse rendering objectives. For example, \citet[Section 5]{gkioulekas2016evaluation} explain the need to use uncorrelated sets of paths for unbiased estimation of squared-loss derivatives; and \citet[Section 4]{Nicolet:2023:CV} explain the bias when using other loss functions. These works also highlight the critical role of unbiasedness in ensuring convergence of stochastic gradient optimization, justifying our own focus on reducing bias while maintaining sample efficiency in reverse-mode derivatives. Given these parallels, our U-statistic estimator could potentially be useful also in differentiable rendering applications.

\section{Implementation and evaluation}\label{sec:evaluation}
We implement \diffbvp{} WoS on top of the open-source \emph{Zombie} library \citep{Sawhney:2023:Zombie}, without any changes to its WoS or WoSt routines for solving \forwardbvp{} BVPs. For optimization, we use a combination of \emph{vector Adam} \citep{ling2022vectoradam} for shape parameters (\eg, mesh vertices), and \emph{Adam} \citep{kingma2015adam} for all others parameters (\eg, pose, rotation, color values). We use a constant learning rate of $10^{-3}$, and exponential annealing of the number of walks per evaluation point (\cref{sec:sgd}).

Before applying our solver to shape optimization problems in \cref{sec:geometric_examples}, in this section we evaluate several aspects of its performance: We first evaluate the accuracy of its derivative estimates against finite differencing (\cref{sec:finite_differences}). We then study the impact of noisy derivatives on optimization (\cref{sec:sgd}). We finally conduct an ablation study of hyperparameters and design choices (\cref{sec:ablation}).

\subsection{Comparison to finite differencing}\label{sec:finite_differences}
\begin{figure}[t]
	\centering
	\includegraphics[width=\columnwidth]{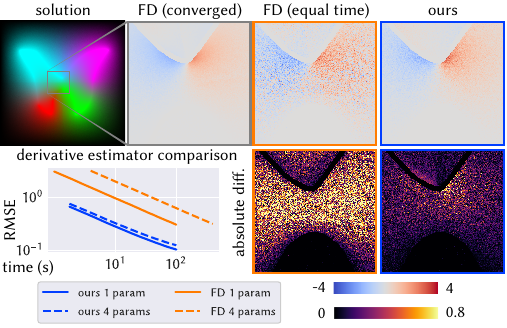}
        \caption{Finite-difference (FD) approximations of derivatives ({\figloc{middle-right}}) require only \forwardbvp{} walks, but scale poorly with parameter count. \Diffbvp{} WoS (\figloc{far-right}) instead computes derivatives for all parameters \(\param\) with a single \diffbvp{} walk, leading to less noisy results at equal time.}
	\label{fig:finite_difference}
\end{figure}

To validate the accuracy of the derivative estimates output by our \diffbvp{} WoS estimator, we compare them in \cref{fig:finite_difference} against estimates from finite differencing (specifically using forward differences). We make two observations:
\begin{enumerate*}
	\item When run till convergence, both techniques produce very similar outputs, thus providing evidence in support of the accuracy of our \diffbvp{} WoS estimator. 
	\item When run for equal time, \diffbvp{} WoS produces estimates of much lower RMSE than finite differencing. 
\end{enumerate*}
The performance improvement \diffbvp{} WoS provides becomes more pronounced as the number of parameters increases. The reason is that, to compute derivatives with respect to $N$ parameters, finite differencing must perform $N+1$ independent \forwardbvp{} walks; by contrast, \diffbvp{} WoS needs to perform only one \diffbvp{} walk irrespective of the value of $N$. The \textit{only} overhead on \diffbvp{} WoS as $N$ increases is storing a sparse vector of derivatives and computing, at the end of the walk, additional normal velocity terms $\normalDispField\paren{\point, \param}$ for any parameter that influences the boundary at the terminal location. Importantly, the number of WoS calls---or equivalently, the total number of walks---is constant with respect to $N$.

\subsection{Optimization with noisy derivatives}\label{sec:sgd}
\begin{figure}[t]
	\centering
	\includegraphics[width=\columnwidth]{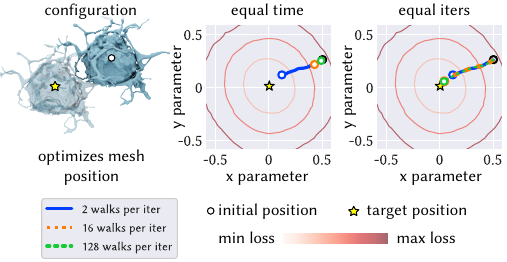}
        \caption{Monte Carlo estimation allow us to trade off between efficiency and noise during optimization by choosing the number of walks used for each derivative estimate. At equal time (\figloc{middle}), noisy derivatives make more progress towards the optimal parameters \(\pi\), while at equal iterations (\figloc{right}), refined derivatives reach a more optimal solution.}
	\label{fig:sgd}
\end{figure}
In \cref{fig:sgd}, we show an experiment where we optimize the position of a mesh so as to match the solution to a screened Poisson equation on a nearby slice plane. We use this setting to evaluate how the variance of derivative estimates impacts optimization convergence and runtime. To this end, we optimize the mesh position by varying the number of \diffbvp{} walks per point (WPP) on the slice plane. Using a smaller WPP results in derivative estimates that are noisier yet also faster to compute; and conversely for larger WPP.

\begin{figure*}[t]
	\centering
	\includegraphics{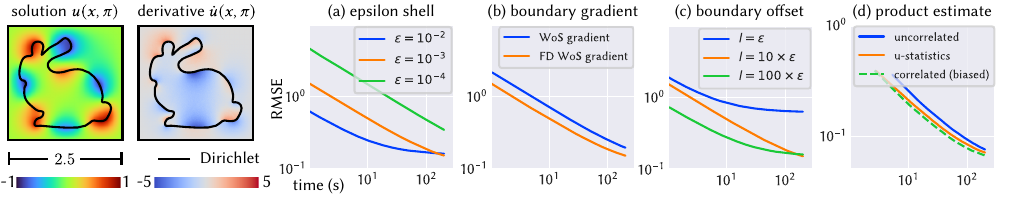}
	\caption{We ablate various hyperparameters of \diffbvp{} WoS on a 2D scene with spatially varying Dirichlet boundary conditions, to study the convergence of derivative estimates. Derivatives are computed with respect to a translation of the mesh in the $y$-direction.}
	\label{fig:ablation}
\end{figure*}

Our experiments show that, at \emph{equal time}, using noisy derivatives (2 WPP) performs much better (\ie, achieve faster progress towards the target mesh position) than using accurate derivatives (16 WPP or 128 WPP). On the other hand, at \emph{equal number of iterations}, using more accurate derivatives converges faster than using noisy derivatives. This behavior highlights a critical advantage of stochastic gradient optimization, namely its ability to improve optimization runtime by using noisy (but unbiased) gradients to trade off between accuracy of descent directions versus iteration cost \citep{StochasticBook,bottou2008tradeoffs}. In practice, to ensure stable convergence, we follow \citet{pfeiffer2018stochastic} and progressively increase WPP using the annealing schedule
\begin{equation}\label{eqn:update_rule}
	\text{WPP}_t = \text{WPP}_0 \cdot \exp(\log(\text{WPP}_T/\text{WPP}_0) \cdot t / N).
\end{equation}
Here $T$ is the number of total iterations, $t = 1,\dots,T$ is the current iteration, and $\text{WPP}_0$ and $\text{WPP}_T$ are the number of walks at the start and end of the optimization (resp.). This schedule provides fast-to-compute, but approximate descent directions in the early stages of the optimization where accuracy is not as important, and more accurate directions in the latter stages.

\subsection{Ablations}\label{sec:ablation}

In \cref{fig:ablation}, we ablate hyperparameters of \diffbvp{} WoS to compare the RMSE of estimated derivatives against reference derivatives computed using finite-differences. We use a 2D bunny mesh normalized to fit inside a unit sphere along with spatially varying Dirichlet boundary conditions, and compute the derivative $\solnDeriv\paren{\point, \param}$ with respect to a $y$-translation of the mesh.

\paragraph{$\varepsilon$-shell size} As \cref{fig:ablation}(a) shows, a larger $\varepsilon$ value leads to lower variance but larger bias. We use $\varepsilon = 10^{-3}$ (corresponding to $\nicefrac{1}{1000}\times$ the dimension of the scene's bounding cube) for all other experiments, as it provides a reasonable balance between bias and variance.

\paragraph{Normal derivative estimator} As \cref{fig:ablation}(b) shows, our backward-difference estimator (\cref{eqn:dudn_fd_estimate}) achieves 20\% lower RMSE than the estimator by \citet{BVC} (\cref{eqn:dudn_mc_estimate}, implemented in Zombie \citep{Sawhney:2023:Zombie} with antithetic sampling and control variates for reduced variance). We thus use the backward-difference estimator in all other experiments. We expect the performance difference between the two estimators to be smaller if the point $y \in \partial \overline{\ball}$ in \cref{eqn:dudn_mc_estimate} is sampled using cosine-weighted sampling \citep[Section 13.6.3]{Pharr:2023:PBR}.

\paragraph{Normal derivative offset} As \cref{fig:ablation}(c) shows, the offset $\offset$ in the backward-difference estimator \labelcref{eqn:dudn_fd_estimate} can result in increased bias if it is either too small or too large. In the extreme case when $\offset < \varepsilon$, WoS terminates immediately and returns the estimate $\hat{\solution}(\point - \offset \cdot \normal_\point) = \dirichlet(\point)$, thus incorrectly always outputting zero normal derivative estimates. At $\offset = \varepsilon$, half of the first WoS steps will be inside the epsilon shell, which still leads to significant bias. Moving in the opposite direction, a large value $\offset = 100\times\varepsilon$ results in noticeable bias due to discretization error. Empirically, we found $\offset = 10\times\varepsilon$ to perform well, and thus use this value for all other experiments.

\paragraph{U-statistic estimator} \cref{fig:ablation}(d) compares our U-statistic estimator (with a batch size $B=8$) and the uncorrelated and correlated estimators from \cref{sec:backward}. Relative to the uncorrelated estimator, the correlated estimator increases bias but reduces variance, overall improving RMSE (which includes both bias and variance) by about $10-25\%$ depending on the WPP.
Our U-statistic estimator achieves comparable RMSE to the correlated estimator, while remaining unbiased. We choose to use the U-statistic estimator in all other experiments, despite the fact that the correlated estimator slightly outperforms it in terms of RMSE. This choice is because prior work in differentiable rendering \citep{Nicolet:2023:CV,10.1111:cgf.14344} has shown that low-variance yet biased derivative estimates can hinder convergence of stochastic gradient optimization.

\section{Geometric examples}\label{sec:geometric_examples}

We apply \diffbvp{} WoS on a range of PDE-constrained shape optimization tasks, including: diffusion-based pose estimation (\cref{sec:pose_estimation}) and surface reconstruction (\cref{sec:diffuse_optical_tomography}), thermal design (\cref{sec:thermal_design}), image-space curve inflation (\cref{sec:curve_inflation}), and inverse diffusion curves (\cref{sec:inverse_diffusion_curves}). In most examples we optimize only boundary geometry, though we show joint optimization of boundary geometry and data in \cref{sec:inverse_diffusion_curves}. We report detailed statistics for each optimization task in \cref{tab:performance}. The project website includes:
\begin{enumerate*}
	\item a video file visualizing the optimization process for all of our examples; and
	\item code for our implementation.
\end{enumerate*}
\begin{figure}[b]
	\centering
	\includegraphics[width=\columnwidth]{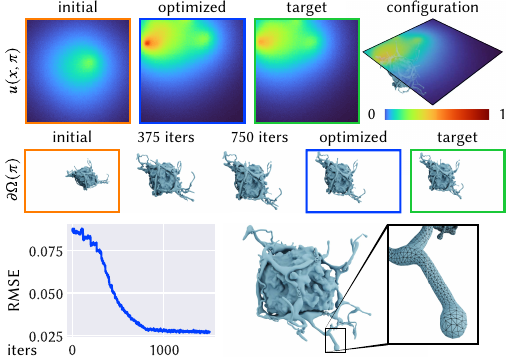}
	\caption{We recover the rotation, position, and scale of a complex boundary mesh (116k faces) to match a solution profile on an image plane (\figloc{top right}).}
	\label{fig:pose}
\end{figure}

\begin{figure*}[t]
	\centering
	\includegraphics{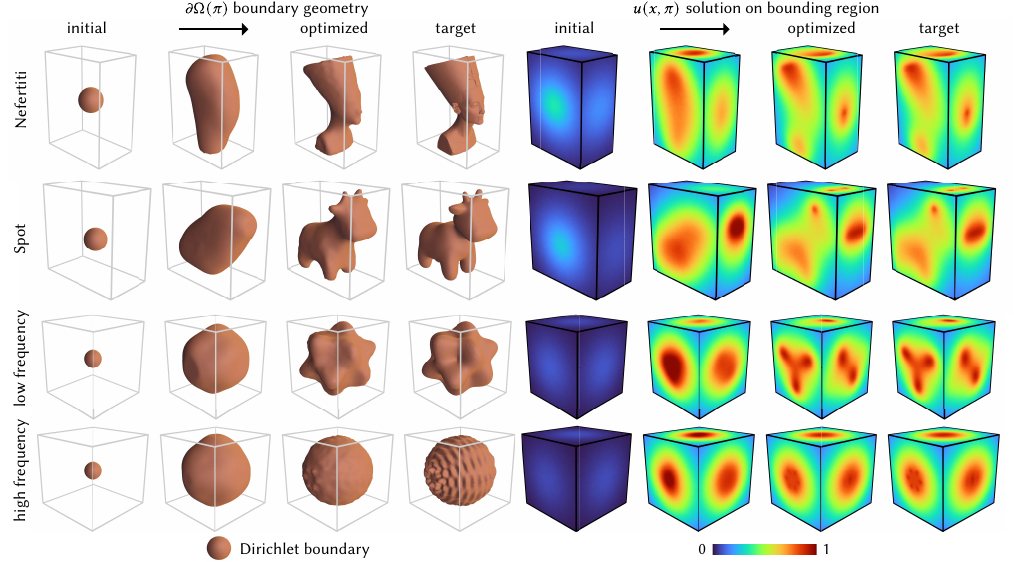}
	\caption{Inspired by shape-from-diffusion applications, we use measurements on a bounding box (\figloc{far right, column}) to evolve 3D models starting from a sphere (\figloc{left}). As with any reconstruction through a diffusive medium, high-frequency surface details have negligible influence on the solution, and are therefore not captured in the reconstructions (\figloc{bottom row}).}
	\label{fig:DOT}
\end{figure*}

\subsection{Pose estimation}\label{sec:pose_estimation}

\paragraph{Background} Pose estimation seeks to recover the position, orientation, and scale of a known shape. A challenging subset of this problem uses measurements of light partially obscured by scattering on a diffusive surface~\citep{Isogawa:2020:PoseEstimation} or through a diffusive medium~\citep{Raviv:2014:PoseEstimation}. We consider the latter scenario in \cref{fig:pose}, where we approximate radiative transport in diffusive media with a screened Poisson equation~\citep[Ch. 5]{Wang:2007:BiomedicalImaging}.

\paragraph{Setup} In \cref{fig:pose}, we model a fixed-shape emissive object as a Dirichlet boundary, and optimize seven pose parameters: four  for rotation and scale (Euler-Rodrigues parameterization), and three for translation. The Dirichlet boundary influences the image plane measurements via the \forwardbvp{} BVP \labelcref{eqn:bvp} with $\source = 0$ and $\dirichlet = 1$.

\paragraph{Output-sensitive optimization} Our \diffbvp{} WoS algorithm can efficiently optimize shape functionals \labelcref{eqn:functional} that are localized, through the masking function $\Mask$, in small subsets of the domain $\domain\paren{\param}$: As a pointwise estimator, \diffbvp{} WoS can estimate gradients only at locations where $\Mask = 1$, unlike conventional grid-based solvers that are constrained to output derivatives over the entire domain---thus wasting significant compute. This output sensitivity of \diffbvp{} WoS makes it well-suited to application settings where we can only measure physical quantities (\eg, light, heat, electric potentials) locally. To demonstrate output sensitivity, in \cref{fig:pose} we optimize pose by minimizing a shape functional that uses the squared loss $\solnLoss$ and a mask $\Mask$ to compare with a reference solution only on an image plane (\cref{fig:pose}, top right corner). \Diffbvp{} WoS allows us to focus all computation only on this image plane.

\subsection{Shape from diffusion}\label{sec:diffuse_optical_tomography}
\paragraph{Background} Differentiable Monte Carlo rendering algorithms \citep{zhang2023projective}, alongside effective preconditioners \citep{Nicolet:2021:LSI}, have recently shown great promise for shape reconstruction tasks from radiometric images, providing benefits such as geometric robustness and scalability, output sensitivity, and trivial parallelism. \Diffbvp{} WoS can bring the same benefits to shape reconstruction tasks from measurements of diffusive phenomena that can be modeled via elliptic PDEs. Such measurements can arise in medical imaging applications due to deep tissue scattering \citep{schweiger:1995:FEMDOT, arridge1999optical}, or from alternative imaging modalities such as thermal imaging \citep{Harbrecht:2013:ShapeFromHeat}. In \cref{fig:teaser,fig:DOT}, we show proof-of-concept experiments for such shape-from-diffusion tasks.

\paragraph{Setup} As in the pose estimation experiment, we model an emissive object of unknown shape as a Dirichlet boundary with constant Dirichlet data. We represent the object as triangular mesh, initialized to a spherical shape. We optimize the mesh vertices to match a reference solution specified on the six faces of a bounding box, as in \cref{fig:teaser} (top right). The emissive object influences the solution on the bounding box via the \forwardbvp{} BVP \labelcref{eqn:bvp} with $\source = 0$ and $\dirichlet = 1$.

\begin{figure*}[t]
	\centering
	\includegraphics{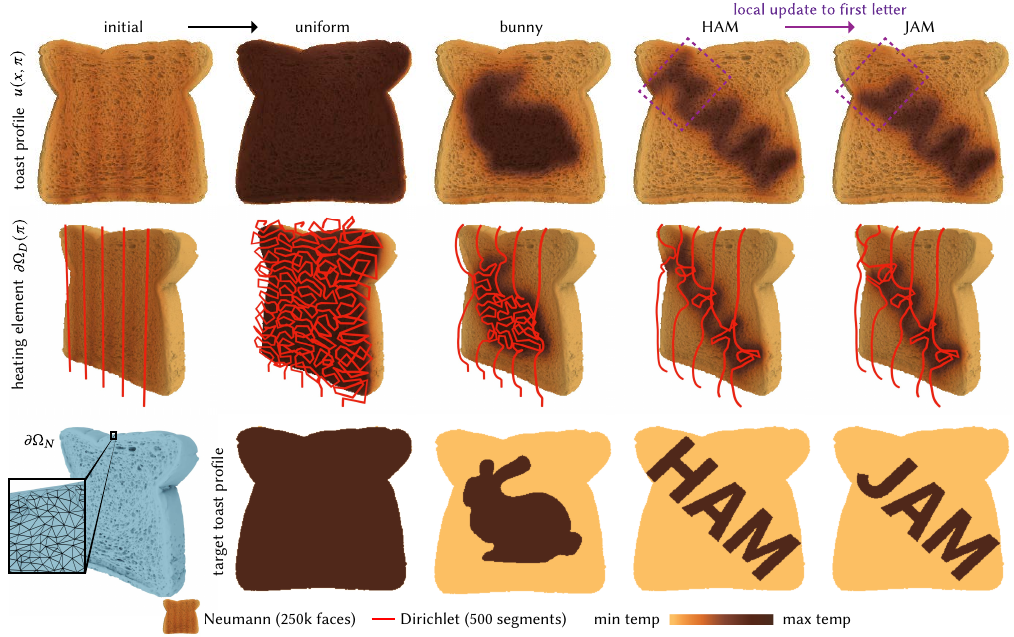}
	\caption{We optimize the heating elements of a toaster (Dirichlet boundary, \figloc{middle row}) to achieve a target temperature profile (\figloc{bottom row}) on a slice of bread (Neumann boundary, \figloc{top row}). For partial updates to a target profile (\eg “ham” to “jam”, \figloc{bottom right}), we leverage locality-sensitive optimization by computing derivatives only over the modified profile.}
	\label{fig:toaster}
\end{figure*}

\paragraph{Gradient preconditioning} During optimization, we use the Laplacian gradient preconditioner from \citet{Nicolet:2021:LSI}, and remesh every 100 iterations~\citep{Botsch:2004:Remesh}. In addition to Vector Adam, we also follow the suggestion of \citet{Nicolet:2021:LSI} to adapt step size using the maximum momentum across all parameters.

\paragraph{High-frequency details} Surface details are faithfully reconstructed to the extent to which they influence diffusive measurements. Due to diffusion, fine geometry details have little influence on the solution at points away from the emissive boundary. We showcase this phenomenon in \cref{fig:DOT} (\figloc{bottom row}): reconstructing a low-frequency approximation to the ground truth high-frequency sphere geometry perfectly reproduces the reference solution.

\subsection{Optimization-driven thermal design}\label{sec:thermal_design}

\paragraph{Background} Elliptic PDEs are at the heart of classic optimization-driven design problems ranging from finding airfoil shapes that minimize drag at cruising speeds, to constructing electromagnets that induce fields with specific characteristics \citep{Pironneau:1984:OptimalShapeDesign}. Recently, thermal systems have become a major focus of optimization-driven design~\citep{Zhan:2008:ThermalDesign, Dbouk:2017:HeatTransfer}, as the shape of components can critically impact, for instance, the performance of integrated circuits or the effectiveness of heat exchangers.

\paragraph{Setup} In \cref{fig:toaster}, we optimize toaster heating elements to achieve a target temperature profile on a slice of bread. Following \citet[Section 6.5]{WoSt}, we model conductive heat transfer from the toaster wires (corresponding to a parameterized Dirichlet boundary $\boundaryDirichlet\paren{\param}$ with 500 line segments) to the slice of bread (corresponding to a fixed Neumann boundary $\boundaryNeumann$ with 250k faces), by solving the BVP \labelcref{eqn:mixed_bvp} with $\source = 0$, $\dirichlet = 1$, $\robin = 0$, and $\neumann = 0$. As this is a mixed Neumann-Dirichlet BVP, we use the \diffbvp{} WoSt algorithm of \cref{sec:diffwost} for optimization.

Similar to \citet{WoSt}, we visualize the solution using a phenomenological model that maps temperature $\solution\paren{\point, \param}$ to color values. We further constrain the heating elements by regularizing their polyline geometry with a bending and stretching energy~\citep{Bergou:2008:DER}. Though we cannot achieve arbitrarily sharp temperature profiles due to the physical constraints of diffusive heat transfer, we find the closest physically realizable profiles. 

\paragraph{Output-sensitive optimization} The heating elements are optimized with respect to a temperature loss defined on the Neumann boundary $\boundaryNeumann$. To determine the reference temperature on $\boundaryNeumann$, we project a target temperature profile $\solution_{\mathrm{ref}}$ from an image plane $\mathcal{S}$ onto $\boundaryNeumann$ via ray tracing---we denote this projection operation as $\projection\paren{\point}$. Doing so, we can avoid directly integrating our loss over all 250k faces on $\boundaryNeumann$, and instead integrate it over just the image plane,
\begin{align}
	\shapeObjective\paren{\param} \equiv
	\int_{\mathcal{S}} \norm{\solution\paren{\projection\paren{\point}, \param} - \solution_{\mathrm{ref}}\paren{\point}}^2 \ud \point
\end{align}
This ``deferred shading'' approach evaluates derivatives at the minimum resolution necessary to capture the details in the target temperature profile (\ie, $128^2$ resolution grid), and only from visible locations on $\boundaryNeumann$ (\ie, the front of the slice of bread). Compared to a naive evaluation at ~124k mesh faces of $\boundaryNeumann$, deferred shading reduces the number of evaluation points by an order of magnitude.

\paragraph{Local modifications} When updating a temperature profile resembling the word ``ham'' to the word ``jam'', we initialize with the heating elements optimized for ``ham''. Output-sensitive optimization lets us optimize with respect to the loss only over the letter ``j'', which reduces average iteration time by over 80\%.

\subsection{Image-space curve inflation}\label{sec:curve_inflation}
\begin{figure}[t]
	\centering
	\includegraphics[width=\columnwidth]{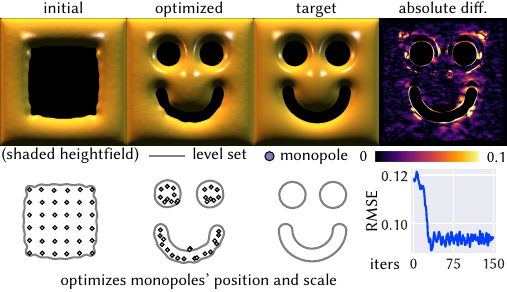}
	\caption{\citet{Baran:2009:NIC} define smooth heightfields over a domain as the solution to a Poisson equation with zero Dirichlet boundary conditions. We optimize an implicit boundary, specifically the position and scale of harmonic radial basis functions, to achieve a target heightfield visualized as a shaded surface (\figloc{top row}).}
	\label{fig:implicit}
\end{figure}
\paragraph{Background} Many prior works have observed the central relationship between silhouettes and 3D shape~\citep{Ikeuchi:1981:ShapeFromShading, Cipolla:1998:VisualMotionCurves, DeCarlo:2003:SuggestiveContours}, which allows depth to be modeled in image space as the solution to a Poisson equation with boundary conditions imposed along silhouettes~\citep{Ikeuchi:1981:ShapeFromShading, Gorelick:2006:PoissonSilhouettes}. Inspired by this relationship, \citet{Baran:2009:NIC} propose an artistic \emph{curve inflation} model where height is modeled as a solution to the BVP \cref{eqn:bvp} with $\screening = 0$, $\source = 4$, and $\dirichlet = 0$.
Intuitively, the height $\solution$ is zero on the boundary, and is smoothly ``inflated'' into the interior of $\shape\paren{\param}$. Thus controlling the boundary of the BVP allows controlling the resulting height field. 
We consider the inverse problem in \cref{fig:implicit}, where we recover the boundary from a heightfield produced with curve inflation. 

\paragraph{Setup} In \cref{fig:implicit}, we optimize the boundary $\boundary\paren{\param}$ to match a reference heightfield $u_{\mathrm{ref}}\paren{\point}$---which we visualize as a shaded surface. We represent the boundary $\boundary\paren{\param}$ \emph{implicitly} as the zero-level set of a function $h\paren{\point, \param}$ parameterized as the sum of $N$ harmonic monopoles---each with a scale $a_n\in\mathbb{R}$ and position $p_n\in\mathbb{R}^2$
\begin{equation}
	\!\!\!h\paren{\point, \param} \equiv c + \sum_{n=1}^N \frac{a_n}{\norm{\point - p_n}},\; \boundary\paren{\param} \equiv \curly{\point\in\R^2 : h(\point, \param) = 0}.
\end{equation}
By using an implicit representation, we can easily handle large topological changes to the boundary, as we show in \cref{fig:implicit}. Applying \diffbvp{} WoS in this setting requires simply implementing conservative closest point queries for this implicit representation. As $h$ is a \emph{harmonic function}, we can achieve this using \emph{Harnack bounds} \citep{Gillespie:2024:RTH}: these bounds provide a conservative radius for an empty sphere at any point, as well as a gradient-sensitive condition $|h\paren{\point}| < |\nabla h\paren{\point}| \varepsilon$ for terminating walks at an $\varepsilon$-shell.

\paragraph{Implicit boundary integrals} Direct sampling of the boundary integral in the shape functional derivative \labelcref{eqn:functional_deriv} is not feasible in this example, because the boundary is defined implicitly. We instead approximate the boundary integral as a volume integral,
\begin{align}
\frac{\dd \shapeObjective}{\dd \param}\paren{\param} &\approx \int_{\domain\paren{\param} \backslash \boundary_\varepsilon\paren{\param}} \solnDeriv\paren{\point, \param}\solnLossDeriv(\solution\paren{\point, \param}) \ud \point \nonumber \\
	&+ \frac{1}{2\varepsilon}\int_{\boundary_\varepsilon\paren{\param}} \normalDispField\paren{\point, \param} \solnLoss(\solution\paren{\point, \param}) \ud \point,
\end{align}
where $\boundary_\varepsilon\paren{\param} \equiv \{\point\in\domain\paren{\param} : \min_{\pointalt\in\boundary\paren{\param}}\|\point-\pointalt\| < \varepsilon \}$. In the limit of $\epsilon\to 0$ we recover the original boundary integral.

\subsection{Inverse diffusion curves}\label{sec:inverse_diffusion_curves}
\begin{figure}[t]
	\centering
	\includegraphics[width=\columnwidth]{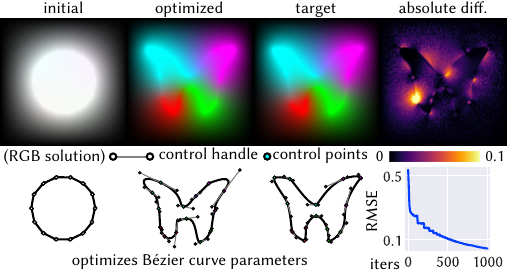}
	\caption{Inspired by \citet{zhao2017inverse}, we optimize B\'{e}zier curves and RGB boundary conditions to generate a diffusion solution that matches a target image. Unique to \diffbvp{} WoS, we can directly optimize the control points and handles of the B\'{e}zier curves.}
	\label{fig:bezier}
\end{figure}

\begin{table*}[t]
    \centering
    \caption{\label{tab:performance} We report the average iteration time of \diffbvp{} WoS across a variety of applications. For certain applications, we increase the number of \diffbvp{} walks per point on an exponential schedule (indicated by ``initial $\to$ final'' in the table). The number of \diffbvp{} walks lists the walks per iteration, whereas the number of \forwardbvp{} walks lists the number of recursive walks to estimate the \diffbvp{} boundary condition, for each \diffbvp{} walk. All applications were run on a 12 core i9-10920X Intel CPU, except for curve inflation which was prototyped on a NVIDIA 3090 GPU.}
	\begin{tabular*}{\linewidth}{@{\extracolsep{\fill}} *{8}{c} @{}}
        application & geometry \ $\boundary\paren{\param}$ & parameters \ $\param$ & \# eval pts &\# diff. walks & \# \forwardbvp{} walks & \# iters &avg. iter time (s)\\
        \midrule
        \hyperref[sec:pose_estimation]{pose estimation} & mesh (58k vertices) & pose matrix & $128^2$ & 2 $\to$ 64 & 2 & 1.5k& 8s\\
        \hyperref[sec:diffuse_optical_tomography]{surface reconstruction}  & mesh (2.5k vertices) & vertex pos.& $6\times 64^2$ &16 $\to$ 256 & 16 & 1k & 82s\\
		\hyperref[sec:thermal_design]{thermal design} & polylines (500 vertices) & vertex pos.& $128^2$ &4 $\to$ 16 & 8 & 250 & 12s\\
		\hyperref[sec:curve_inflation]{curve inflation} & implicit (32 monopoles) & pos. \& scale & $128^2$ &4 & 4 & 150 & 2s\\
		\hyperref[sec:inverse_diffusion_curves]{diffusion curves} & B\'{e}ziers (14 curves) & pos. \& tangents & $128^2$ & 2 $\to$ 64 & 2 & 1k & 3s\\
    \end{tabular*}
\end{table*}

\paragraph{Background} \citet{Orzan:2013:DiffusionCurves} introduced \emph{diffusion curves} as a vector graphic primitive that models images as the solution $\solution\paren{\point}: \shape\to [0, 1]^3$ to a Laplace equation---that is, the BVP \cref{eqn:bvp} with $\screening = 0$ and $\source = 0$---with RGB-valued Dirichlet boundary conditions $\dirichlet\paren{\cdot,\param}$ along a boundary $\boundary\paren{\param}$ represented as B\'{e}zier curves. 
Rather than manually create diffusion curves, \citet{zhao2017inverse} optimize $\boundary\paren{\param}$ and $\dirichlet(\cdot, \param)$ to match a target image. They use FEM to solve \cref{eqn:bvp}, which constrains them to use a polyline boundary representation. As we show in \cref{fig:bezier}, with \diffbvp{} WoS we can instead optimize a boundary encoded using B\'{e}zier curves.

\paragraph{Setup} We jointly optimize over both the B\'{e}zier curves $\boundary\paren{\param}$ and the RGB-valued Dirichlet boundary conditions $\dirichlet\paren{\cdot, \param}$ to match a target image. We enforce $G^1$ continuity connections between the curves, by parameterizing the boundary so that adjacent B\'{e}ziers have co-linear control handles. Our optimization also accounts for continuous Dirichlet conditions arising from linear interpolation of RGB values between control points.

\paragraph{Monte Carlo length regularization} We follow \citet[Section 5.3]{zhao2017inverse} and use a length regularizer $R\paren{\param}\equiv\alpha\int_{\boundary\paren{\param}} \ud l\paren{\point}$ to penalize geometrically complex curves. During optimization, we use Monte Carlo estimation of the derivative of this regularizer:
\begin{equation}
	\dot{R}\paren{\boundary\paren{\param}} =  \alpha \int_{\boundary\paren{\param}} \kappa\paren{\point} \normalDispField\paren{\point}\ud l\paren{ \point}
\end{equation}
where $\kappa$ is the curvature of the B\'{e}zier curve (available in closed form). To ensure the regularization is scaled appropriately relative to estimator variance, we decay its strength using a schedule $\alpha_t \coloneqq \alpha_0 \paren{\nicefrac{\mathrm{WPP}_0}{\mathrm{WPP}_t}}^{-\nicefrac{1}{2}}$, where $\alpha_t$ and $\mathrm{WPP}_t$ are the regularization strength and WPP (resp.) at the $t^{th}$ gradient iteration. We choose the decay rate of the schedule to match the expected Monte Carlo convergence rate of $\bigO(n^{-1/2})$.

\section{Conclusion and Future Work}\label{sec:discussion}
We introduced a \diffbvp{} walk on spheres method to estimate derivatives with respect to perturbations of the domain boundary. In developing this method, we crucially take the approach of differentiating the functional relationship in \cref{eqn:bvp} rather than the estimator in \cref{eqn:wos}. This approach allows us to easily adapt the the classic WoS method, used for solving \forwardbvp{} PDEs \citep{Muller:1956:WOS, Sawhney:2020:MCG}, into a method for solving a set of nested BVPs for the parameter derivative. 

Even though these nested BVPs are well known in the shape optimization literature \citep{henrot2018shape}, using WoS to estimate them brings to PDE-constrained shape optimization the unique benefits of Monte Carlo methods---output-sensitive evaluation, geometric robustness and flexibility, and the scalability of stochastic gradient-based optimization. We illustrated these benefits through varied geometric examples, yet our method is only a first step towards applying Monte Carlo PDE solvers to real-world inverse problems. To help further this goal, we discuss generalizations and improvements for enhancing the capabilities of our method.

\paragraph{More sophisticated physical models} We focused on the screened Poisson equation due to its conceptual simplicity and widespread use in graphics and other sciences. This PDE already places our technique within scope for inverse problems used in practice. For instance, medical imaging with electrical impedance tomography \citep{cheney1999electrical} involves optimizing a variable diffusion coefficient in a screened Poisson equation with Robin boundary conditions \citep{uhlmann2009electrical}---recent work generalizes WoS to PDEs of this form \citep{svWoS,WoStRobin}. Thermal imaging techniques like shape from heat \citep{Chen:2015:ReconstructionFromIR} require simulating both the propagation of infrared light and the transfer of heat through solids via conduction. Already \citet{bati2023coupling} have developed such a Monte Carlo solver for coupled heat transfer; a \diffbvp{} variant could be derived by considering both differentiable rendering and \diffbvp{} WoS. However, more general inverse problems require extending our technique to more general PDEs. For example, the design of heat exchangers \citep{feppon2019shape} involves simulating the coupled physics of convection and conduction, often via simplified models for the flow of fluids and heat diffusion (resp.)---simulation of fluids in a Monte Carlo framework with WoS (and related solvers) has recently garnered attention in graphics \citep{Rioux:2022:MCFluid,Jain:2024:MCFluid,sugimoto2024velocity}.

\paragraph{Parameterized Robin boundaries} We discussed in \cref{sec:diffwost} how WoSt \citep{WoSt, WoStRobin} can be used, in a limited capacity, to optimize the Dirichlet boundary in a mixed Dirichlet-Robin BVP. Optimizing parameterized Robin boundaries $\boundaryNeumann\paren{\param}$ is likewise possible by evaluating a corresponding nested Dirichlet-Robin BVP \citep[Page 230, Eq 5.79]{henrot2018shape}. In this case, nesting occurs via a \diffbvp{} Robin boundary condition evaluated at each step of the \diffbvp{} WoSt estimator. However unlike \cref{eqn:mixed_bvp_deriv}, this boundary condition involves evaluating second-order spatial gradients that can lead to high variance---lowering noise of such gradients is necessary to make this approach practical.

\paragraph{Open boundaries} In \cref{sec:optimization} we restricted our discussion to closed boundaries, and this restriction is a current limitation of our method. The reason for this restriction is that, when working with an open boundary, the \diffbvp{} BVP will include a boundary condition with a Dirac delta term on the boundary perimeter. WoS cannot importance sample such localized boundary conditions, and thus this limitation also constrains \diffbvp{} WoS. This limitation could be overcome by using alternative bidirectional solvers \citep{qi2022bidirectional,WoStRobin} that simulate \emph{reverse} random walks starting from the boundary. Support for open boundaries should be possible by using reverse \diffbvp{} walks with our method.

\paragraph{Variance reduction} More generally, any improvement in estimation quality to the underlying WoS estimator should translate directly to \diffbvp{} WoS---this includes recently developed variance reduction strategies for WoS \citep{qi2022bidirectional,BVC,bakbouk2023mean,li2023neural} we did not consider in this paper for simplicity. Specific to geometric optimization, it is also worthwhile investigating how to reuse derivative estimates with similar values across successive iterations in an optimization to improve efficiency; such strategies have found success in differentiable Monte Carlo rendering, in the form of recursive control variates~\citep{Nicolet:2023:CV} and reservoir-based temporal importance resampling~\citep{Chang:2023:ReSTIRDiffRender}. Finally, we should improve upon the basic finite difference scheme we use in \cref{sec:diffwos} to lower both the variance and bias in our normal derivative estimates---the alternative normal derivative estimator developed by \citet{Yu:2024:DiffPoisson} applies directly to our method as well.

\begin{acks}
	This work was supported by National Science Foundation (NSF) awards 2008123 and 2212290; National Institute of Food and Agriculture award 2023-67021-39073; a gift from Adobe Research; NSF Graduate Research Fellowship DGE2140739 and an NVIDIA graduate fellowship for Bailey Miller; a Packard Foundation Fellowship for Keenan Crane; and Alfred P. Sloan Research Fellowship FG202013153 for Ioannis Gkioulekas. Rohan Sawhney thanks Ken Museth for supporting this work and the authors thank Gilles Daviet for helpful discussion.
\end{acks}

\bibliographystyle{ACM-Reference-Format}
\bibliography{WoStDiff}

\appendix

\section{Differentiation of Dirichlet Boundary Condition with Texture Mapping}\label{app:proof}

We consider boundary data $\dirichlet \coloneqq \dirichletReference \circ \invEmbedding$ that composes the inverse parameterization $\invEmbedding\paren{\cdot, \param}: \boundary\paren{\param} \to \bracket{0,1}^2$ with a texture function $\dirichletReference\paren{\cdot, \param}: \bracket{0,1}^2 \to \R$. We determine the boundary condition to use in the \diffbvp{} BVP by implicitly differentiating the boundary condition of the BVP \labelcref{eqn:bvp}---which we reproduce here:
\begin{align}\label{eqn:boundary_condition_texture}
	\solution\paren{\point, \param} = \dirichlet\paren{\point, \param} = \dirichletReference\paren{\invEmbedding\paren{\point, \param}, \param} \quad \text{ on } \boundary\paren{\param}.
\end{align}
Differentiating the left-hand side and using the chain rule yields:
\begin{equation}\label{eqn:deriv1}
	\frac{\dd}{\dd \param} \solution\paren{\point, \param} = \paramderiv{\solution}\paren{\point, \param} + \nabla \solution\paren{\point, \param} \cdot \dispField\paren{\point, \param}.
\end{equation}
Doing likewise for the right-hand side gives us:
\begin{align}
	\frac{\dd}{\dd\param} \dirichletReference\paren{\invEmbedding\paren{\point, \param}, \param} &= \dirichletReferenceDeriv\paren{{\invEmbedding\paren{\point, \param}, \param}} \nonumber \\
	&+ \nabla \dirichletReference\paren{\invEmbedding\paren{\point, \param},\param} \frac{\dd}{\dd \param} \invEmbedding\paren{\point, \param} \\
	&= \dirichletReferenceDeriv\paren{{\invEmbedding\paren{\point, \param}, \param}} \nonumber \\
	&+ \nabla \dirichletReference\paren{\invEmbedding\paren{\point, \param},\param} \nonumber \\
	&\quad\quad\quad\cdot\paren{\paramderiv{\invEmbedding}\paren{\point, \param} + \nabla \invEmbedding\paren{\point, \param} \dispField\paren{\point, \param}}.
\end{align}
We use the identity $\nabla\dirichletReference \cdot \nabla \invEmbedding = \nabla \dirichlet$ and make dependence on $\param$ implicit to simplify this expression:
\begin{align}\label{eqn:deriv2}
	\frac{\dd}{\dd\param} \dirichletReference\paren{\invEmbedding\paren{\point}} &= \dirichletReferenceDeriv\paren{{\invEmbedding\paren{\point}}} \nonumber \\
	&+ \nabla \dirichletReference\paren{\invEmbedding\paren{\point}} \cdot \paramderiv{\invEmbedding}\paren{\point} + \nabla \dirichlet\paren{\point} \dispField\paren{\point, \param}.
\end{align}
Lastly, we note that, as \cref{eqn:boundary_condition_texture} requires that $\solution$ and $\dirichlet$ agree along the boundary, their tangential gradients are the same. Thus, by equating \cref{eqn:deriv1,eqn:deriv2} and eliminating tangential components on both sides, we arrive at the boundary condition:
\begin{align}\label{eqn:boundary_deriv_mapped_rep}
	\solnDeriv\paren{\point} &= \dirichletReferenceDeriv\paren{{\invEmbedding\paren{\point}}} + 
	\nabla \dirichletReference\paren{\invEmbedding\paren{\point}} \cdot 
	\paramderiv{\invEmbedding}\paren{\point} \nonumber \\
	&+ \paren{\frac{\partial \dirichlet}{\partial \normal}\paren{\point} - \frac{\partial \solution}{\partial \normal}\paren{\point}}\normalDispField\paren{\point, \param} \quad \text{ on } \boundary\paren{\param},
\end{align}
which is the same as \cref{eqn:boundary_deriv_mapped}.

\section{The adjoint boundary value problem}\label{app:adjoint}

We briefly comment on an alternative formulation for computing derivatives of PDE-constrained shape functionals $\shapeObjective\paren{\param}$ as in \cref{eqn:functional}. \citet{zhao2017inverse} and \citet[Section 5.8]{henrot2018shape} show that we can express the derivative of the shape functional as:
\begin{equation}\label{eqn:functional_deriv_adjoint}
	\frac{\dd \shapeObjective}{\dd \param}\paren{\param} = 
		\int_{\boundary\paren{\param}} \frac{\partial \adjoint}{\partial \normal}\paren{\pointalt} \paren{\frac{\partial \dirichlet}{\partial \normal}\paren{\pointalt} - \frac{\partial \solution}{\partial \normal}\paren{\pointalt}} \normalDispField\paren{\pointalt, \param} \ud \sigma\paren{\pointalt},
\end{equation}
where $\adjoint\paren{\cdot, \param}: \domain\paren{\param} \to \R$ is the solution to the \emph{adjoint boundary value problem}:
\begin{equation}
	\label{eqn:bvp_adjoint}
	\begin{array}{rcll}
		\Delta \adjoint\paren{\point} - \screening \adjoint\paren{\point} &=& \solnLossDeriv(\solution\paren{\point})	& \text{ in } \shape\paren{\param}, \\
		\adjoint\paren{\point} &=& 0 & \text{ on } \boundary\paren{\param}. \\
	\end{array}
\end{equation}
Compared to \cref{eqn:functional_deriv}, \cref{eqn:functional_deriv_adjoint} is simpler because it requires estimating only a boundary integral and no domain integral. Compared to the \diffbvp{} BVP \labelcref{eqn:bvp_deriv}, the adjoint BVP is also a screened Poisson equation with Dirichlet boundary conditions; however, the \forwardbvp{} BVP is nested within the source term, and not the boundary data. Consequently, though the solution $\adjoint$ could still be estimated using WoS, nesting is more difficult: As the source term is invoked at every recursion step of WoS, nesting would require launching a walk for estimating $\solution$ at \emph{every step} along the walk for estimating $\adjoint$, resulting in quadratic complexity. By contrast, nesting for the \diffbvp{} BVP requires launching a walk for estimating $\solution$ at \emph{only the last step} of the walk for estimating $\solnDeriv$, maintaining linear complexity (\cref{alg:wos_deriv}). This complexity difference motivated our choice to use the \diffbvp{} BVP. The quadratic complexity of the adjoint approach could potentially be overcome using off-centered connections (\citep[Section 5.2]{svWoS} and \citep[Section 4.3]{Sawhney:2020:MCG}) to merge the walks for $\solution$ back into the walk for $\adjoint$; or caching schemes \citep{BVC,li2023neural} that estimate $\solution$ without a walk. Such approaches are exciting future research directions.

\section{More general shape functional}\label{app:shape_functional}

In some of our experiments in \cref{sec:geometric_examples}, we use a more general shape functional of the form:
\begin{align}
	\shapeObjective\paren{\param} &\equiv \int_{\domain\paren{\param}} \Mask\paren{\point}\solnLoss\paren{\solution\paren{\point,\param}} \ud \point \nonumber \\
	&+ \int_{\boundary\paren{\param}} \mask\paren{\pointalt}\solnLossBoundary\paren{\solution\paren{\pointalt,\param}} \ud \mathrm{A}\paren{\pointalt}, \label{eqn:general_functional}
\end{align}
where $\solnLoss, \solnLossBoundary: \R \to \R$ are differentiable loss functions, and $\Mask, \mask: \R^3 \to \curly{0,1}$ are binary mask functions. Compared to \cref{eqn:functional}, the shape functional of \cref{eqn:general_functional} includes an additional loss term evaluated on the boundary $\boundary\paren{\param}$. Differentiating this shape functional produces \citep[p. 239]{henrot2018shape}:
\begin{align}\label{eqn:general_functional_deriv}
	\frac{\dd \shapeObjective}{\dd \param}\paren{\param} &= \int_{\domain\paren{\param}} \Mask\paren{\point}\ \solnDeriv\paren{\point, \param}\solnLossDeriv(\solution\paren{\point, \param}) \ud \point \nonumber \\
	&+ \int_{\boundary\paren{\param}} \mask\paren{\pointalt}\ \solnDeriv\paren{\pointalt, \param}\solnLossBoundaryDeriv(\solution\paren{\pointalt, \param}) \nonumber \\
	&\quad\quad+ \normalDispField\paren{\pointalt, \param}\Big(\Mask\paren{\pointalt}\solnLoss(\solution\paren{\pointalt, \param}) \nonumber \\
	&\quad\quad\quad\quad\quad\quad+ \mask\paren{\pointalt}\curvature\paren{\pointalt}\solnLossBoundary(\solution\paren{\pointalt, \param}) \nonumber \\
	&\quad\quad\quad\quad\quad\quad+ \mask\paren{\pointalt}\solnLossBoundaryDeriv(\solution\paren{\pointalt, \param})\frac{\partial \solution}{\partial \normal}\paren{\pointalt, \param}\Big) \ud \mathrm{A}\paren{\pointalt},
\end{align}
where $\solnLossDeriv, \solnLossBoundaryDeriv$ are the derivatives of the scalar losses $\solnLoss, \solnLossBoundary$; and $\curvature$ is the mean curvature. \Cref{eqn:general_functional_deriv} is valid under the same two assumptions we made for \cref{eqn:functional_deriv}, and requires computing values of $\solution$, $\nicefrac{\partial\solution}{\partial\normal}$, and $\solnDeriv$ like that equation. Thus we can estimate the shape functional derivative of \cref{eqn:general_functional_deriv} using the same algorithm we developed in \cref{sec:derivatives} for the derivative of \cref{eqn:functional_deriv}.

\end{document}